\documentclass[12pt]{article}
\usepackage{enumerate}
\usepackage{amsfonts}
\usepackage{amsmath}
\usepackage{amssymb}
\usepackage{amsthm}

\def\H{\mathcal{H}}

\def\S{\mathfrak{S}}

\def\T{\mathfrak{T}}
\def\B{\mathfrak{B}}

\newcommand{\rank}{\mathrm{rank}}
\newcommand{\id}{\mathrm{Id}}
\newcommand{\Tr}{\mathrm{Tr}}

\newcommand{\shs}{\hspace{1pt}}

\newcounter{defin}  \newcounter{lemma}  \newcounter{theorem}
\newcounter{property} \newcounter{corol}  \newcounter{remark} \newcounter{example}

\newenvironment{lemma}{\par\refstepcounter{lemma}
     \textbf{Lemma \thelemma.} }{\rm\par}
\newenvironment{theorem}{\par\refstepcounter{theorem}
     \textbf{Theorem \thetheorem.}\ }{\rm\par}
\newenvironment{property}{\par\refstepcounter{property}
     \textbf{Proposition \theproperty.}\ }{\rm\par}
\newenvironment{corollary}{\par\refstepcounter{corol}
     \textbf{Corollary \thecorol.} }{\rm\par}

\newenvironment{remark}{\par\refstepcounter{remark}
     \textbf{Remark \theremark.}}{\rm\par}

\begin{document}

\title{On discontinuity of information characteristics of quantum systems and channels}
\author{M.E. Shirokov\footnote{Steklov Mathematical Institute, RAS, Moscow, email:msh@mi.ras.ru}}
\date{}
\maketitle

\vspace{-10pt}

\begin{abstract}
Quantitative analysis of discontinuity of basic characteristics of quantum states and channels
is presented.

First we consider general estimates for discontinuity jump (loss) of the von Neumann entropy for a given converging sequence of states. It is shown, in particular, that for any sequence the loss of entropy is upper bounded by the loss of mean energy
(with the coefficient  characterizing Hamiltonian of a system).

Then we prove that
discontinuity jumps of several correlation and entanglement measures in composite quantum systems
are upper bounded by loss of one of the marginal entropies (with a corresponding coefficient).

We also analyse discontinuity of the output entropy of a quantum operation and of basic information charateristics of a quantum channel with respect to simultaneous variations of an input state and of a channel.
\end{abstract}

\vspace{-10pt}

\tableofcontents

\section{Introduction}

One of the main difficulties in study of infinite-dimensional quantum systems consists in discontinuity of basic characteristics of quantum states and channels (such as von Neumann entropy, conditional entropy, quantum mutual information, entanglement measures, etc.).
This shows  necessity to find conditions for local continuity of such characteristic. The first results in this direction seems to be Simon's convergence theorems for the von Neumann entropy \cite[the Appendix]{Simon}. Since then many different continuity conditions for the entropy and other basic information quantities have been found (see \cite{H-SCI,W,Winter,CMI} and the references therein).

In this paper we present quantitative analysis of discontinuity of several important characteristics of quantum states and channels starting with the von Neumann entropy.

In Section 3 we consider general estimates  for discontinuity jumps of the von Neumann entropy and an expression for these jumps based on the approximating technique (Propositions 1,3 and their corollaries). We also consider relations between discontinuity jumps of the entropy and majorization  (Proposition 2). Then we focus attention on estimating discontinuity of the entropy on the set of states with bounded mean energy, i.e. states $\rho$ satisfying the inequality
\begin{equation}\label{f-ineq}
  \Tr H\rho\leq E
\end{equation}
where $H$ is the Hamiltonian of a system. It is well known that the entropy is continuous on this set if (and only if)
$\Tr\shs e^{-\lambda H}$ is finite for all $\lambda>0$ \cite{W}. Explicit continuity bounds for the entropy on this set were recently obtained by  Winter \cite{Winter}.
We analyse discontinuity jumps (losses) of the entropy  on the set determined by inequality (\ref{f-ineq}) in the case of logarithmic growth of the eigenvalues of $H$, i.e. when
\begin{equation}\label{b-cond}
 \Tr\shs e^{-\lambda H}<+\infty\; \textrm{ for some }\;\lambda>0.
\end{equation}
It is shown that for any converging sequence of states the loss of entropy is upper bounded by the loss of mean energy
with the coefficient $g(H)$ -- the infimum of all $\lambda$ in (\ref{b-cond}) (Proposition 4). \smallskip

In Section 4  we show that
discontinuity jumps of several measures of classical and quantum correlations in composite quantum systems
are upper bounded by
discontinuity jump of one of the marginal entropies (with a corresponding  coefficient).
The main conclusion obtained by joining these results and the observation from Section 3 can be briefly formulated as follows: \emph{if Hamiltonians of quantum subsystems satisfy condition (\ref{b-cond}) then discontinuity of many charateristics of a composite quantum state is related to the loss of mean energy in one of the subsystems.}\footnote{There exist correlation measures whose discontinuity \emph{is not related} to discontinuity of the entropy (and hence to the loss of mean energy) \cite{Chen&Co,W&Co}.}\smallskip

In Section 5  we analyse discontinuity of the output entropy of a quantum operation and of the basic information charateristics of a quantum channel: the constrained Holevo capacity, the quantum mutual information and the coherent information. We obtain estimates for discontinuity jumps of these characteristics with respect to simultaneous variations of an input state and of a channel expressed via loss of the input (output) entropy.

\section{Preliminaries}

Let $\mathcal{H}$ be a separable Hilbert space,
$\mathfrak{B}(\mathcal{H})$ and $\mathfrak{T}( \mathcal{H})$ --
Banach spaces of all bounded operators and of all trace-class
operators in $\mathcal{H}$, $\mathfrak{T}_{+}(\mathcal{H})$ -- the
cone of positive operators in $\mathfrak{T}( \mathcal{H})$,
$\mathfrak{S}(\mathcal{H})$ -- the set of quantum states (operators
in $\mathfrak{T}_{+}(\mathcal{H})$ with unit trace)
\cite{H-SCI,N&Ch}.

Denote by $I_{\mathcal{H}}$ the identity operator in a Hilbert space
$\mathcal{H}$ and by $\id_{\mathcal{\H}}$ the identity
transformation of the Banach space $\mathfrak{T}(\mathcal{H})$.

A \emph{quantum operation} $\,\Phi$ from a system $A$ to a system
$B$ is a completely positive trace non-increasing linear map
$\mathfrak{T}(\mathcal{H}_A)\rightarrow\mathfrak{T}(\mathcal{H}_B)$,
where $\mathcal{H}_A$ and $\mathcal{H}_B$ are Hilbert spaces
associated with the systems $A$ and $B$. A trace preserving quantum
operation is called \emph{quantum channel} \cite{H-SCI,N&Ch}.

For any  quantum channel $\,\Phi:A\rightarrow B\,$ Stinespring's theorem implies the existence of a Hilbert space
$\mathcal{H}_E$ and of an isometry
$V:\mathcal{H}_A\rightarrow\mathcal{H}_B\otimes\mathcal{H}_E$ such
that
\begin{equation}\label{St-rep}
\Phi(\rho)=\mathrm{Tr}_{E}V\rho V^{*},\quad
\rho\in\mathfrak{T}(\mathcal{H}_A).
\end{equation}
The minimal dimension of $\H_E$ is called \emph{Choi rank} of $\shs\Phi$. The quantum  channel
\begin{equation}\label{c-channel}
\mathfrak{T}(\mathcal{H}_A)\ni
\rho\mapsto\widehat{\Phi}(\rho)=\mathrm{Tr}_{B}V\rho
V^{*}\in\mathfrak{T}(\mathcal{H}_E)
\end{equation}
is called \emph{complementary} to the channel $\Phi$
\cite[Ch.6]{H-SCI}.

The \emph{quantum relative entropy} for two operators $\rho$ and
$\sigma$ in $\mathfrak{T}_{+}(\mathcal{H})$ is defined as follows
(cf.\cite{L-2})
\begin{equation}\label{re-def}
H(\rho\,\|\shs\sigma)=\sum_{i=1}^{+\infty}\langle
i|\,\rho\log\rho-\rho\log\sigma+\sigma-\rho\,|i\rangle,
\end{equation}
where $\{|i\rangle\}_{i=1}^{+\infty}$ is the orthonormal basis of
eigenvectors of the operator $\rho$ and it is assumed that
$H(\rho\,\|\sigma)=+\infty$ if $\,\mathrm{supp}\rho$ is not
contained in $\mathrm{supp}\shs\sigma$. This definition implies
$H(\lambda\rho\,\|\shs\lambda\sigma)=\lambda
H(\rho\,\|\shs\sigma)$ for $\lambda\geq0$. \smallskip

We will use the following result of the purification
theory \cite{H-SCI}.\smallskip
\begin{lemma}\label{p-lemma}
\textit{Let $\mathcal{H}$ and $\mathcal{K}$ be Hilbert spaces such
that $\,\dim\mathcal{H}=\dim\mathcal{K}$. For an arbitrary pure
state $\omega_{0}$ in
$\,\mathfrak{S}(\mathcal{H}\otimes\mathcal{K})$ and an arbitrary
sequence $\{\rho_{k}\}$ of states in $\,\mathfrak{S}(\mathcal{H})$
converging to the state
$\rho_{0}=\mathrm{Tr}_{\mathcal{K}}\omega_{0}$ there exists a
sequence $\{\omega_{k}\}$ of pure states in
$\,\mathfrak{S}(\mathcal{H}\otimes\mathcal{K})$ converging to the
state $\omega_{0}$ such that\break
$\rho_{k}=\mathrm{Tr}_{\mathcal{K}}\omega_{k}$ for all $\,k$.}
\end{lemma}\smallskip

We will repeatedly use the following simple lemmas in which $X$ is an arbitrary metric space.\smallskip

\begin{lemma}\label{vsl+} \emph{Let $f$, $g$ and $h$ be functions on $X$ such that $f+g=h$ and $\{x_n\}$ a sequence converging
to $\,x_0$ such that $f(x_0),g(x_0)$ and $h(x_0)$ are finite.}

\smallskip

\noindent\emph{If the function $g$ is lower semicontinuous then
$$
\limsup_{n\rightarrow\infty}f(x_n)-f(x_0)\leq \limsup_{n\rightarrow\infty}h(x_n)-h(x_0)
$$
If the function $h$ is lower semicontinuous then
$$
f(x_0)-\liminf_{n\rightarrow\infty}f(x_n)\leq \limsup_{n\rightarrow\infty}g(x_n)-g(x_0).
$$}

\end{lemma}\smallskip

\emph{Proof.} Since $\,f(x_n)+g(x_n)=h(x_n)\,$ for all $\,n$, we have
$$
\limsup_{n\rightarrow\infty}f(x_n)+\liminf_{n\rightarrow\infty}g(x_n)\leq \limsup_{n\rightarrow\infty}h(x_n).
$$
By subscribing the equality $\,f(x_0)+g(x_0)=h(x_0)\,$ from this inequality and by using the lower semicontinuity of $\,g\,$  we obtain the first assertion of the lemma. The second assertion follows from the first one with $f'=-f, h'=g, g'=h$. $\square$

 \smallskip

\begin{lemma}\label{app-l} \emph{Let $\{f_k\}$ and $\{g_k\}$ be nondecreasing sequences of continuous functions on $X$ pointwise converging respectively to functions $f$ and $g$. If $f(x)-f_k(x)\leq g(x)-g_k(x)$ for all $\,x\in X$  then
$$
\limsup_{n\rightarrow\infty}f(x_n)-f(x_0)\leq \limsup_{n\rightarrow\infty}g(x_n)-g(x_0)
$$
for any sequence $\{x_n\}$ converging
to a state $x_0$ such that $g(x_0)<+\infty$.}
\end{lemma}\smallskip

\emph{Proof.} It suffices to note that continuity of the functions $f_k$ and $g_k$ imply
 $$
\limsup_{n\rightarrow\infty}h(x_n)-h(x_0)=\lim_{k\rightarrow\infty}\limsup_{n\rightarrow\infty}(h-h_k)(x_n),\quad h=f,g,
$$
provided that $g(x_0)$ (and hence $f(x_0)$) are finite. $\square$ \smallskip

\begin{lemma}\label{vsl} \emph{Let $\,\{f_k\}$ be a non-increasing  sequence of functions on $\,X$ pointwise converging to a function $f$ and $\{x_n\}$  a sequence converging
to $x_0$ such that $f(x_0)<+\infty$. If
$$
\limsup_{n\rightarrow\infty}f_k(x_n)-f_k(x_0)\leq C\;
\textit{ for all } \,k\, \textit{ then }\;
\limsup_{n\rightarrow\infty}f(x_n)-f(x_0)\leq C.
$$}
\end{lemma}\smallskip

\section{On discontinuity  of the von Neumann entropy}

\subsection{General estimates}

The \emph{von Neumann entropy} $H(\rho)=\mathrm{Tr}\eta(\rho)$, where $\eta(x)=-x\log x$, is a basic characteristic of a
state $\rho\in\mathfrak{S}(\mathcal{H})$. It has the homogeneous  extension to the cone
$\mathfrak{T}_{+}(\mathcal{H})$ (cf.\cite{L-2})\footnote{Here and in
what follows $\log$ denotes the natural logarithm.}
\begin{equation}\label{ent-ext}
H(\rho)=[\mathrm{Tr}\rho] H\!\left(\frac{\rho}{\mathrm{Tr}
\rho}\right)=\mathrm{Tr}\eta(\rho)-\eta(\mathrm{Tr}\rho),\quad \rho
\in\mathfrak{T}_{+}(\mathcal{H}).
\end{equation}

This extension naturally arises in applications, for example, in analysis of the output entropy of quantum trace-non-preserving operation (see Sect.5.1). Nonnegativity, concavity and lower semicontinuity of the von Neumann
entropy on the cone $\mathfrak{T}_{+}(\mathcal{H})$ follow from the
corresponding properties of this function on the set
$\mathfrak{S}(\mathcal{H})$ \cite{W, L-2}.

By the lower semicontinuity of the entropy $H(\rho)$ its discontinuity jumps for a given sequence $\{\rho_n\}\in\T_+(\H)$ converging to an operator $\rho_0$ can be characterised by the nonnegative value
$$
\mathrm{dj}\!\left\{H(\rho_n)\right\}\doteq\limsup_{n\rightarrow+\infty}H(\rho_n)-H(\rho_0),
$$
where it is assumed that $\mathrm{dj}\!\left\{H(\rho_n)\right\}=+\infty$ if $H(\rho_0)=+\infty$. This value can be called \emph{the entropy loss}
corresponding to the sequence $\{\rho_n\}$.\footnote{The term "entropy loss" is used in literature in different senses \cite{EL,EL+}.}
\smallskip

We begin with the following simple but useful observation. \smallskip

\begin{property}\label{simple-p}
\emph{Let $\{\rho_n\}\in\T_+(\H)$ be a sequence converging to an operator $\rho_0$. Then
\begin{equation}\label{Gen-UB}
\mathrm{dj}\!\left\{H(\rho_n)\right\}\leq\limsup_{n\rightarrow\infty}\Tr\rho_n(-\log\sigma_n)-\Tr\rho_0(-\log\sigma_0)
\end{equation}
for any sequence $\{\sigma_n\}\in\T_+(\H)$ converging to an operator $\sigma_0$, where it is assumed that
the right hand side is equal to $\shs+\infty\shs$ if $\;\Tr\rho_0(-\log\sigma_0)=+\infty$.}
\end{property}\smallskip

Note that $"="$ trivially holds in (\ref{Gen-UB}) if $\,\sigma_n=\rho_n\,$ for all $n$.\smallskip

\emph{Proof.} It follows from (\ref{re-def}) and (\ref{ent-ext}) that
\begin{equation}\label{u-rel}
H(\rho_n)+H(\rho_n\|\shs\sigma_n)+f(\rho_n)-\Tr\sigma_n=\Tr\rho_n(-\log\sigma_n)
\end{equation}
for all $n\geq0$, where $f(\rho_n)=\eta(\Tr\rho_n)+\Tr\rho_n$. Hence
$$
\limsup_{n\rightarrow\infty}H(\rho_n)+\liminf_{n\rightarrow\infty}H(\rho_n\|\shs\sigma_n)+
\lim_{n\rightarrow\infty}[f(\rho_n)-\Tr\sigma_n]\leq\limsup_{n\rightarrow\infty}\Tr\rho_n(-\log\sigma_n).
$$
By subscribing equality (\ref{u-rel}) with $\,n=0\,$ from this inequality and by using the lower semicontinuity of the relative entropy we obtain
(\ref{Gen-UB}). $\square$ \smallskip

We will use Proposition \ref{simple-p} below. Now consider two simple corollaries.
\smallskip

Let $\{|k\rangle\}$ be an orthonormal basis in $\H$. For any state $\rho\in\S(\H)$ we may consider the probability distribution
$\pi(\rho)=\{\langle k|\rho|k\rangle\}$. It is well known that $H(\rho)\leq S(\pi(\rho))$, where $S$ is the Shannon entropy -- a lower semicontinuous function on the set of all countable probability distributions equipped with the $\ell_1$ metric. Proposition \ref{simple-p} shows that a similar relation hold for jumps of the entropy  corresponding to converging sequences $\{\rho_n\}$ and $\{\pi(\rho_n)\}$.\footnote{It is easy to see that the map $\,\rho\mapsto\pi(\rho)\,$ is continuous.}
\smallskip

\begin{corollary}\label{simple-p-c-1}
\emph{For any sequence  $\{\rho_n\}\in\S(\H)$ converging to a state $\rho_0$ we have
\begin{equation}\label{S-UB}
\mathrm{dj}\!\left\{H(\rho_n)\right\}\leq\mathrm{dj}\!\left\{S(\pi(\rho_n))\right\}\doteq\limsup_{n\rightarrow\infty}S(\pi(\rho_n))-S(\pi(\rho_0)),
\end{equation}
where it is assumed that
$\,\mathrm{dj}\!\left\{S(\pi(\rho_n))\right\}$ is equal to $\shs+\infty\shs$ if $\,S(\pi(\rho_0))=+\infty$.}
\end{corollary}\smallskip

Note that $"="$ holds in (\ref{S-UB}) if the sequence $\{\rho_n\}$ consists of states diagonalisable in the basis $\{|k\rangle\}$.\smallskip

\emph{Proof.} It suffices to take $\,\sigma_n=\sum_k\langle k|\rho_n|k\rangle|k\rangle\langle k|\,$ for all $\,n\geq0,$ and to apply  Proposition \ref{simple-p}. $\square$
\medskip

By subadditivity of the von Neumann entropy
$H(\omega_{AB})\leq H(\omega_{A})+H(\omega_{B})$
for any bipartite state $\omega_{AB}$, where $\omega_{A}\doteq\Tr_B\shs\omega_{AB}$ and $\omega_{B}\doteq\Tr_A\shs\omega_{AB}$ are marginal states.
Similar relation holds for jumps of the entropy. \smallskip

\begin{corollary}\label{simple-p-c-3} \emph{Let  $\{\omega_{AB}^n\}$ be a sequence of bipartite states converging to a state $\omega_{AB}^0$. Then}
\begin{equation*}
\mathrm{dj}\!\left\{H(\omega_{AB}^n)\right\}\leq \mathrm{dj}\!\left\{H(\omega_{A}^n)\right\}+\mathrm{dj}\!\left\{H(\omega_{B}^n)\right\}.
\end{equation*}
\end{corollary}\smallskip

\emph{Proof.} It suffices to take $\,\sigma_n=\omega_{A}^n\otimes\omega_{B}^n\,$ for all $\,n\geq0\,$ and to apply Proposition \ref{simple-p}. $\square$
\smallskip

The triangle inequalities
$H(\omega_{X})\leq H(\omega_{AB})+H(\omega_{Y})$, $XY=AB,BA$  and the implication $\,H(\omega_{AB})=0\;\Rightarrow\;H(\omega_{A})=H(\omega_{B})\,$ have the following $\mathrm{dj}$-versions
\begin{equation*}
\mathrm{dj}\!\left\{H(\omega_{X}^n)\right\}\leq\mathrm{dj}\!\left\{H(\omega_{AB}^n)\right\}+2\mathrm{dj}\!\left\{H(\omega_{Y}^n)\right\},\quad XY=AB,BA,
\end{equation*}
where the factor $2$ can be removed if $\,\{H(\omega_{Y}^n)\}$ is a converging sequence, and
\begin{equation*}
\mathrm{dj}\!\left\{H(\omega_{AB}^n)\right\}=0\quad\Rightarrow\quad\mathrm{dj}\!\left\{H(\omega_{A}^n)\right\}=\mathrm{dj}\!\left\{H(\omega_{B}^n)\right\},
\end{equation*}
valid for any sequence $\{\omega_{AB}^n\}$  converging to a state $\omega_{AB}^0$.
These relations directly follow from Theorem \ref{DB-OE} and Corollary 11 in Section 5 (where $\Phi_n=\Phi$ is a partial trace). The last implication  means that  \emph{continuity of the bipartite entropy implies coincidence of the marginal entropy losses}.\smallskip

The inequalities
$\,H(\omega_{X})\leq H(\omega_{AB})$, $X=A,B$, for a separable state $\omega_{AB}$  also have $\mathrm{dj}$-versions (see Corollary \ref{simple-p-c-3+}B in the next subsection 3.2).

\subsection{The entropy loss and majorization}

Important role in quantum information theory is played by the special partial order between quantum states called \emph{majorization} (see \cite{Nielsen+,E&M,G&Co} and the references therein). We say that a state $\rho$ majorizes a state $\sigma$ and write $\rho\succ\sigma$ if
\begin{equation}\label{m-def}
\sum_{k=1}^n\lambda_k\geq \sum_{k=1}^n\mu_k\quad\textrm{ for all }\; n\in\mathbb{N},
\end{equation}
where $\{\lambda_k\}$ and $\{\mu_k\}$ are sequences of eigenvalues of $\rho$ and $\sigma$ taken in decreasing order. Denote these sequences respectively by $\rho^\downarrow$ and $\sigma^\downarrow$. \smallskip

It is well known that $\rho\succ\sigma$ implies $H(\rho)\leq H(\sigma)$ \cite{Nielsen+,E&M}. To prove the analogous implication for jumps of the entropy 
we will use the following\smallskip

\begin{property}\label{m-lemma+}
\emph{Let $D$ be the classical relative entropy (the Kullback-Leibler distance). If $\,\rho\shs\succ\sigma$ then
\begin{equation}\label{dH}
H(\sigma)=H(\rho)+D\!\left(\rho^\downarrow\|\sigma^\downarrow\right)+f(\rho,\sigma),
\end{equation}
where $\,f(\rho,\sigma)\,$ is a nonnegative lower semicontinuous function on the closed subset\footnote{It is assumed that the set $\S_\succ$ is  equipped with the Cartesian product topology.} $\,\S_\succ\doteq\{\shs(\rho, \sigma)\,|\,\rho\shs\succ\sigma\}$ of $\,[\S(\H)]^{\times2}$ well defined for states $\rho$ and $\sigma$ with finite entropy by the expression
$$
f(\rho,\sigma)=\Tr(\sigma^\downarrow-\rho^\downarrow)(-\log\sigma^\downarrow)=\sum_{k=1}^{+\infty}(\mu_k-\lambda_k)(-\log\mu_k).
$$}
\end{property}\smallskip

\begin{remark}\label{m-lemma-r} Proposition  \ref{m-lemma+} and Pinsker's inequality show that
$$
\rho\shs\succ\sigma \quad\Rightarrow\quad H(\sigma)-H(\rho)\geq D\!\left(\rho^{\downarrow}\|\sigma^{\downarrow}\right)\geq \textstyle{\frac{1}{2}}\|\sigma^\downarrow-\rho^\downarrow\|_1^2.
$$
This gives a simple proof of the strict monotonicity of the entropy with respect to  majorization (cf. \cite{E&M}).
\end{remark}\smallskip

By Proposition \ref{m-lemma+} for any sequences $\{\rho_n\}$ and $\{\sigma_n\}$  converging respectively to states $\rho_0$ and $\sigma_0$ such that $\rho_n\succ\sigma_n$ for all $\,n$ we have
\begin{equation*}
  \liminf_{n\rightarrow\infty}f(\rho_n,\sigma_n)\geq f(\rho_0,\sigma_0).
\end{equation*}
This and the lower semicontinuity of the function  $(\rho,\sigma)\mapsto D\!\left(\rho^{\downarrow}\|\sigma^{\downarrow}\right)$
make possible to derive from (\ref{dH}) the following observation.\smallskip

\begin{corollary}\label{simple-p-c-2}
\emph{Let $\{\rho_n\}$ and $\{\sigma_n\}$ be sequences of states converging respectively to states $\rho_0$ and $\sigma_0$ such that $\rho_n\succ\sigma_n$ for all $\,n$. Then
\begin{equation*}
\mathrm{dj}\!\left\{H(\rho_n)\right\}\leq\mathrm{dj}\!\left\{H(\sigma_n)\right\}-\Delta_1-\Delta_2\leq\mathrm{dj}\!\left\{H(\sigma_n)\right\},
\end{equation*}
where}
$$
\!\Delta_1=\liminf_{n\rightarrow\infty}D\!\left(\rho_n^{\downarrow}\|\sigma_n^{\downarrow}\right)-
D\!\left(\rho_0^{\downarrow}\|\sigma_0^{\downarrow}\right)\geq 0,\;\; \Delta_2=\liminf_{n\rightarrow\infty}f(\rho_n,\sigma_n)-f(\rho_0,\sigma_0)\geq0.
$$
\end{corollary}\smallskip

\emph{Proof of Proposition \ref{m-lemma+}.} If $H(\sigma)<+\infty$ then
$$
H(\sigma)=H(\rho)+D\!\left(\rho^\downarrow\|\sigma^\downarrow\right)+\sum_{k=1}^{+\infty}(\mu_k-\lambda_k)(-\log\mu_k),
$$
where the last series is nonnegative by the below Lemma \ref{m-lemma}. Lemma \ref{m-lemma} also shows that this series is a limit of the nondecreasing sequence of nonnegative numbers
$$
f_n(\rho,\sigma)=\sum_{k=1}^{+\infty}(\mu_k-\lambda_k)h^n_k,\quad\text{где}\quad\,h^n_k=\min\{n, -\log\mu_k\}.
$$
The function $f_n(\rho,\sigma)$ is continuous on $\S_\succ$ for each  $n$ by Mirsky's inequality $\|\rho^{\shs\downarrow}-\sigma^{\shs\downarrow}\|_1\leq\|\rho-\sigma\|_1$ \cite{M}. So, the function  $f(\rho,\sigma)\doteq \sup_n f_{n}(\rho,\sigma)$ possesses all the properties stated in the proposition . It suffices only to verify that
if $H(\sigma)=+\infty$ but $H(\rho)<+\infty$ and $D\!\left(\rho^\downarrow\|\sigma^\downarrow\right)<+\infty$ then $f(\rho,\sigma)=+\infty$  $\square$.
\smallskip

\begin{lemma}\label{m-lemma} \emph{Let $\,\{\lambda_k\}$ and $\,\{\mu_k\}$ be probability distributions such that
$\,\{\lambda_k\}\succ\{\mu_k\}$.  Then $\,\sum_{k=1}^{+\infty}\lambda_kh_k\leq
\sum_{k=1}^{+\infty}\mu_k h_k\,$ for any nondecreasing sequence $\,\{h_k\}$ of nonnegative numbers.}
\end{lemma}
\medskip

\emph{Proof.} It suffices to note that $\sum_{k=1}^{+\infty}\nu_k h_k=\sum_{k=1}^{+\infty}d_kS^\nu_k+h_1$, $\nu=\lambda,\mu$,
where $\,d_k=h_{k+1}-h_k\geq0\,$ and $\,S^\nu_n=\sum_{k>n}\nu_k$, and to use (\ref{m-def}). $\square$\smallskip

Let $\S_\mathrm{s}(\H_{AB})$ be the set of all separable states in $\S(\H_{AB})$ (defined as the convex closure of all product states in $\S(\H_{AB})$). Theorem 11.0.1 in \cite{Nielsen+} states that
\begin{equation}\label{s-m-r}
\omega_{A}\succ\omega_{AB} \quad\textup{and}\quad \omega_{B}\succ\omega_{AB}
\end{equation}
for any state $\omega_{AB}\in\S_\mathrm{s}(\H_{AB})$ provided $\H_A$ and $\H_B$ are finite-dimensional spaces. To generalize this theorem to the case
$\,\dim\H_A=\dim\H_B=+\infty\,$ it suffices to approximate a separable state $\omega_{AB}$ by any sequence $\{\omega^n_{AB}\}$ of separable states with finite rank marginal states $\omega^n_A$ and $\omega^n_B$ and to note that $\omega^n_{X}\succ\omega^n_{AB}$ for all $n$ implies $\,\omega_{X}\succ\omega_{AB}$, $X=A,B$.\footnote{This approximation is necessary because of the existence of countably nondecomposable separable states in infinite-dimensional bipartite system.}\smallskip

Relation (\ref{s-m-r}), Proposition \ref{m-lemma+} and Corollary \ref{simple-p-c-2} imply the following \smallskip
\begin{corollary}\label{simple-p-c-3+}\emph{ Let $\,\S^\mathrm{f}_\mathrm{s}(\H_{AB})=\left\{\omega_{AB}\in\S_\mathrm{s}(\H_{AB})\,|\,H(\omega_{AB})<+\infty\right\}$. }\smallskip

A) \emph{The functions
$\;\omega_{AB}\mapsto H(\omega_{AB})-H(\omega_{X}),\; X=A,B,$
are nonnegative and lower semicontinuous on the set $\,\S^\mathrm{f}_\mathrm{s}(\H_{AB})$.}\smallskip

B) \emph{If $\,\{\omega^n_{AB}\}$ is a sequence of separable states converging to a state $\,\omega^0_{AB}$ then}
\begin{equation*}
\mathrm{dj}\!\left\{H(\omega_{X}^n)\right\}\leq\mathrm{dj}\!\left\{H(\omega_{AB}^n)\right\},\quad X=A,B.
\end{equation*}
\end{corollary}\smallskip

Corollaries \ref{simple-p-c-3} and \ref{simple-p-c-3+}B show that
\begin{equation*}
\max\!\left\{\mathrm{dj}\!\left\{H(\omega_{A}^n)\right\},\mathrm{dj}\!\left\{H(\omega_{B}^n)\right\}\right\}
\leq\mathrm{dj}\!\left\{H(\omega_{AB}^n)\right\}
\leq\mathrm{dj}\!\left\{H(\omega_{A}^n)\right\}+\mathrm{dj}\!\left\{H(\omega_{B}^n)\right\}
\end{equation*}
for any sequence $\,\{\omega^n_{AB}\}$ of separable states converging to a state $\,\omega^0_{AB}$.\smallskip

Corollary \ref{simple-p-c-3+}A makes possible to prove  lower semicontinuity of the coherent information and of the entropy gain for all infinite-dimensional quantum channels complementary to entanglement-breaking channels.  Following \cite{R&Co} we will call such channels \emph{pseudo-diagonal}.\smallskip

\begin{corollary}\label{c-i-ls} \emph{If $\,\Phi:A\rightarrow B\,$ is a pseudo-diagonal quantum channel then the coherent information
$\,I_c(\Phi,\rho)\doteq H(\Phi(\rho))-H(\widehat{\Phi}(\rho))\,$ and the entropy gain
$\,EG(\Phi,\rho)\doteq H(\Phi(\rho))-H(\rho)\,$ are nonnegative lower semicontinuous functions on the set $\,\left\{\rho\in\S(\H_A)\,|\,H(\Phi(\rho))<+\infty\right\}$.}
\end{corollary}\smallskip

\emph{Proof.} If $\rho_{AR}$ is any purification of an input state $\rho_A$ then
$$
I_c(\Phi,\rho)\doteq H\!\left(\widehat{\Phi}\otimes\id_R(\rho_{AR})\right)-H(\widehat{\Phi}(\rho_A))$$
and
$$
EG(\Phi,\rho)\doteq H\!\left(\widehat{\Phi}\otimes\id_R(\rho_{AR})\right)-H(\rho_A).
$$
Since $\widehat{\Phi}$ is an entanglement-breaking channel, $\widehat{\Phi}\otimes\id_R(\rho_{AR})$ is a separable state in $\S(\H_{ER})$.
So, the assertions of the corollary follow from Corollary \ref{simple-p-c-3+}A and Lemma \ref{p-lemma}. $\square$

\subsection{Use of the approximating technique}

In \cite{SSP} it is shown  that the function $\,\T_+(\H)\ni\rho\mapsto H(\rho)\,$ is a pointwise limit of the nondecresing sequence of concave continuous functions
\begin{equation}\label{H-k-def}
H_{k}(\rho)\doteq\sup_{\{\pi_{i},\rho_{i}\}\in\mathcal{P}_{k}(\rho)}
\sum_{i}\pi_{i}H(\rho_{i}),\qquad \rho\in
\mathfrak{T}_{+}(\mathcal{H}),
\end{equation}
where $\mathcal{P}_{k}(\rho)$ is the sets of all countable ensembles of  positive trace
class operators of rank $\leq k$  with the average state $\rho$ (if $\rho$ is a state then the supremum in (\ref{H-k-def}) can be taken over all countable ensembles of \emph{states} of rank $\leq k$ with the average state $\rho$).

The function $H_{k}$ may be called \textit{$k$-approximator} of the
von Neumann entropy. For any
$\rho\in\mathfrak{T}_{+}(\mathcal{H})$ the difference
$\,\Delta^H_{k}(\rho)=H(\rho)-H_{k}(\rho)\,$ between the von Neumann
entropy and its $k$-approximator can be expressed as follows
\begin{equation}\label{Delta}
\Delta^H_{k}(\rho)=\inf_{\{\pi_{i},\rho_{i}\}\in\mathcal{P}_{k}(\rho)}
\sum_{i}\pi_{i}H(\rho_{i}\|\rho),
\end{equation}
where $H(\cdot\|\cdot)$ is the extended quantum relative entropy defined by
(\ref{re-def}). \smallskip

The sequence $\{H_k\}$ is used in \cite{SSP} for analysis of continuity of the von Neumann entropy. It can be also used for
estimating  discontinuity jumps of the entropy. Since the sequence $\{H_k\}$ pointwise converges to the function $H$ and consists of continuous functions, expression (\ref{Delta}) implies the following

\smallskip

\begin{property}\label{appr-DB}
\emph{Let $\{\rho_n\}\subset\T_+(\H)$ be a sequence converging to an operator $\rho_0$ with finite $H(\rho_0)$. Then $\,\mathrm{dj}\!\left\{H(\rho_n)\right\}=\limsup_{n\rightarrow\infty}\Delta^H_{k}(\rho_n)-\Delta^H_{k}(\rho_0)$ for any $k$ and hence}
\begin{equation*}
\mathrm{dj}\!\left\{H(\rho_n)\right\}=\lim_{k\rightarrow\infty}\limsup_{n\rightarrow\infty}\Delta^H_{k}(\rho_n)=\inf_k\limsup_{n\rightarrow\infty}\Delta^H_{k}(\rho_n).
\end{equation*}
\end{property}\smallskip

Applicability of Proposition \ref{appr-DB} is based on special properties of the function $\Delta^H_{k}$ presented in Lemma 8 in \cite{SSP}.
These properties are derived by using  representation (\ref{Delta}) from the well known analytical properties of the quantum relative entropy.
For example, the joint convexity of the relative entropy implies
\begin{equation}\label{delta-1}
\Delta^H_{k+l}(\rho+\sigma)\leq\Delta^H_{k}(\rho)+\Delta^H_{l}(\sigma),\quad \rho,\sigma\in\T_+(\H),
\end{equation}
while the monotonicity of the relative entropy shows that
\begin{equation}\label{delta-2}
\rho\leq\sigma\quad \Rightarrow \quad\Delta^H_{k}(\rho)\leq\Delta^H_{k}(\sigma),\quad \rho,\sigma\in\T_+(\H),
\end{equation}
where $"\leq"$ in the left side denotes the operator order, and that
\begin{equation}\label{delta-3}
\Delta^H_{mk}(\Phi(\rho))\leq\Delta^H_{k}(\rho)
\end{equation}
for any $\rho\in\T_+(\H)$ and any quantum operation $\Phi$ with Choi rank $\leq m$.\smallskip

By Proposition \ref{appr-DB} properties (\ref{delta-1}) and (\ref{delta-2}) imply the following\smallskip

\begin{corollary}\label{appr-DB+}
\emph{Let $\{\rho_n\}$ and $\{\sigma_n\}$  be sequences of operators in $\T_+(\H)$ converging respectively to operators $\rho_0$ and $\sigma_0$. Then} \begin{equation*}
\max\left\{\mathrm{dj}\!\left\{H(\rho_n)\right\},\mathrm{dj}\!\left\{H(\sigma_n)\right\}\right\}\leq\mathrm{dj}\!\left\{H(\rho_n+\sigma_n)\right\}
\leq\mathrm{dj}\!\left\{H(\rho_n)\right\}+\mathrm{dj}\!\left\{H(\sigma_n)\right\}.
\end{equation*}
\end{corollary}\smallskip

A strengthened version of Corollary \ref{appr-DB+} is obtained in Section 4.2 (Cor.\ref{Chi-DB+}).
\smallskip

Proposition \ref{appr-DB} and  property (\ref{delta-3}) show that the loss of entropy does not increase under action of quantum operations with bounded Choi rank. \smallskip

\begin{corollary}\label{appr-DB++}
\emph{Let $\{\rho_n\}\subset\T_+(\H)$ be a sequence converging to an operator $\rho_0$ and
$\{\Phi_n\}$ a sequence of quantum operations with bounded Choi rank such that the sequence $\{\Phi_n(\rho_n)\}$  converges to the operator $\Phi_0(\rho_0)$. Then}
\begin{equation}\label{C-UB}
\mathrm{dj}\!\left\{H(\Phi_n(\rho_n))\right\}\leq\mathrm{dj}\!\left\{H(\rho_n)\right\}.
\end{equation}
\end{corollary}\smallskip

In particular, (\ref{C-UB}) holds if $\{\Phi_n\}$ is a sequence of quantum operations with bounded Choi rank strongly converging to the operation
$\Phi_0$ (see Sect.5.2).

\subsection{States with bounded energy}

Let $H$ be a positive unbounded operator in a Hilbert space $\H$ which will be treated as a Hamiltonian of a quantum system associated with the space $\H$. Then
$$
\mathcal{K}_{H,E}\doteq\left\{\rho\in\S(\H)\,|\,\Tr H\rho\leq E\right\}
$$
is the set of states with mean energy not exceeding $E$ (here
$\Tr H\rho$ is defined as a limit of the nondecreasing sequence $\{\Tr P_n H\rho\}$ of positive numbers, where $P_n$ is the spectral projector of $H$ corresponding to the interval $[0,n]$).

It is well known (cf.\cite{W}) that the von Neumann entropy is continuous on the set $\mathcal{K}_{H,E}$ if
\begin{equation}\label{Gh}
 \Tr\shs e^{-\lambda H}<+\infty\;\; \textrm{ for all }\; \lambda>0.
\end{equation}
Recently Winter obtained explicit continuity bounds for the von Neumann entropy on the set $\mathcal{K}_{H,E}$ in this case \cite{Winter}. In fact, (\ref{Gh}) is a necessary and sufficient condition of continuity of the entropy on the set
$\mathcal{K}_{H,E}$ if $E$ is greater than the minimal energy level of $H$ (this follows from Proposition \ref{H-DB} below). So, dealing with Hamiltonians not satisfying condition (\ref{Gh}) we have to take into account discontinuity jumps of  the von Neumann entropy (and of the related quantities) on the set $\mathcal{K}_{H,E}$.\smallskip

Introduce the parameter
$$
 g(H)\doteq \inf\{\lambda>0\,|\,\Tr\shs e^{-\lambda H}<+\infty \},
$$
which is assumed to be $+\infty$  if $\,\Tr\shs e^{-\lambda H}=+\infty\,$ for all $\lambda>0$.

If $g(H)<+\infty$ then the operator $H$ has discrete spectrum of finite multiplicity, i.e. it can be represented as follows
\begin{equation}\label{H-form}
  H=\sum_{k=0}^{+\infty}E_k|k\rangle\langle k|,\quad E_{k}\leq E_{k+1},
\end{equation}
where $\{|k\rangle\}$ is an orthonormal basic and $\{E_k\}$ is a nondecreasing sequence of eigenvalues (energy levels) of $H$. Since $\,g(H)=\limsup_k E_{k}^{-1}\log
k$, the inequality  $\,0<g(H)<+\infty\,$ means the logarithmic growth of the sequence $\{E_k\}$.
For any $\lambda>g(H)$ one can introduce the state
$\,\sigma_{\lambda}=[\Tr\shs e^{-\lambda H}]^{-1}e^{-\lambda H}$. Then we have the identity
\begin{equation}\label{l-rel}
H(\rho)+H(\rho\shs\|\shs\sigma_{\lambda})=\lambda \Tr H\rho+C, \quad C=\log[\mathrm{Tr}\shs e^{-\lambda H}],
\end{equation}
valid for any state $\rho$,  which shows that the entropy is bounded on the set $\mathcal{K}_{H,E}$ for any $E>0$.

If $g(H)=+\infty$ then the entropy is not bounded (and hence is not finite\footnote{It is easy to show that finiteness of the entropy on the closed convex set guarantees its boundedness on this set (see the proof of Theorem 1 in \cite{OE}).}) on the set $\mathcal{K}_{H,E}$ if
$\,E>\inf_{\|\varphi\|=1}\langle\varphi|H|\varphi\rangle$.\footnote{This can be shown by noting first that boundedness of the entropy on the set $\mathcal{K}_{H,E}$ implies that $H$ has discrete spectrum of finite multiplicity and then by using the sequence of states (\ref{sequence}) from the proof of Proposition \ref{H-DB} below.}

Assume that the Hamiltonian $H$ has form (\ref{H-form}). For any state $\rho$ introduce its rearrangement $\rho^{\shs\downarrow}$ corresponding to the basic $\{|k\rangle\}$ as (cf. \cite{G&Co})
$$
\rho^{\shs\downarrow}=\sum_{k=0}^{+\infty}\lambda_k|k\rangle\langle k|,
$$
where $\{\lambda_k\}$ is the sequence of eigenvalues of $\rho$ taken in decreasing order. It is clear that $H(\rho^{\shs\downarrow})=H(\rho)$. By using Ky Fan's Maximum Principle  it is easy to show (see the proof of Lemma IV.9 in \cite{G&Co}) that
\begin{equation}\label{rear-in}
  \Tr H\rho^{\shs\downarrow}\leq \Tr H\rho.
\end{equation}
Mirsky's inequality implies $\|\rho^{\shs\downarrow}-\sigma^{\shs\downarrow}\|_1\leq\|\rho-\sigma\|_1$ \cite{M}. So, the map $\rho\mapsto\rho^{\shs\downarrow}$
is continuous.  Hence the functions $\,\rho\mapsto E_H(\rho)\doteq\Tr H\rho\,$ and $\,\rho\mapsto E_H(\rho^{\shs\downarrow})$ are lower semicontinuous on $\S(\H)$ and for a given sequence $\{\rho_n\}\subset\mathcal{K}_{H,E}$ converging to a state $\rho_0$ their discontinuity jumps are characterised by the nonnegative values
$$
\mathrm{dj}\!\left\{E_H(\rho_n)\right\}\doteq\limsup_{n\rightarrow+\infty}E_H(\rho_n)-E_H(\rho_0)\leq E
$$
and
$$
\mathrm{dj}\!\left\{E_H(\rho^{\shs\downarrow}_n)\right\}\doteq\limsup_{n\rightarrow+\infty}E_H(\rho^{\shs\downarrow}_n)-E_H(\rho^{\shs\downarrow}_0)\leq E,
$$
where the last inequality follows from (\ref{rear-in}).\smallskip

\begin{property}\label{H-DB} \emph{Let $\,H$ be a positive operator and  $\,E>E_0\doteq\displaystyle\inf_{\|\varphi\|=1}\langle\varphi|H|\varphi\rangle$.}

\noindent A) \emph{If $g(H)<+\infty$  then
\begin{equation}\label{H-DB-ub}
\mathrm{dj}\!\left\{H(\rho_n)\right\}\leq g(H)\shs\mathrm{dj}\!\left\{E_H(\rho^{\shs\downarrow}_n)\right\}
\leq g(H)\shs\mathrm{dj}\!\left\{E_H(\rho_n)\right\}\leq g(H)(E-E_{0})
\end{equation}
for any converging sequence $\{\rho_n\}\subset\mathcal{K}_{H,E}$. All the bounds in (\ref{H-DB-ub}) are sharp.}\smallskip

\noindent B) \emph{In general case
$$
\sup_{\{\rho_n\}\subset\mathcal{K}_{H,E}}\mathrm{dj}\!\left\{H(\rho_n)\right\}=g(H)(E-E_{0})\leq+\infty,
$$
where the supremum is over all converging sequences $\{\rho_n\}\subset\mathcal{K}_{H,E}$.}
\end{property}\smallskip

\begin{remark}\label{H-DB-r}
The first two inequalities in (\ref{H-DB-ub}) are valid for arbitrary converging sequence $\{\rho_n\}$ if we assume that $\mathrm{dj}\!\left\{E_H(\rho_n)\right\}=+\infty$ in the case $E_H(\rho_0)=+\infty$.
\end{remark}\smallskip

\emph{Proof.} A) Assume that the operator $H$ has form (\ref{H-form}). Since $\,H(\rho^{\shs\downarrow}_n)=H(\rho_n)\,$ for all $\,n$, equality (\ref{l-rel}) and the lower semicontinuity of the relative entropy imply, by Lemma \ref{vsl+}, validity of the first inequality in (\ref{H-DB-ub}) with $g(H)$ replaced by any $\lambda>g(H)$. \smallskip

To prove the second one it suffices, by Lemma \ref{vsl+}, to show that the nonnegative function $f(\rho)\doteq E_H(\rho)-E_H(\rho^{\shs\downarrow})$ is lower semicontinuous on
$\mathcal{K}_{H,E}$. Let
$$
f_m(\rho)\doteq \Tr H_m(\rho-\rho^{\shs\downarrow}),\quad \textrm{where}\quad H_m=\sum_{k=0}^{m-1}E_k|k\rangle\langle k|+E_m\sum_{k=m}^{+\infty}|k\rangle\langle k|.
$$
Since $H_m$ is a bounded operator and the map $\,\rho\mapsto\rho^{\shs\downarrow}$ is continuous, the function $f_m$ is continuous on $\S(\H)$ for all $m$. Since $D_m\doteq H-H_m$ is an operator of the form (\ref{H-form}), inequality (\ref{rear-in}) holds with $H$ replaced  by $D_m$ and hence
$$
f(\rho)-f_m(\rho)=\Tr D_m(\rho-\rho^{\shs\downarrow})\geq0
$$
for any state $\rho$. It is easy to see that $f_m(\rho)$ tends to $f(\rho)$ for any state $\rho$ with finite $E_H(\rho)$. Thus, the function $f$ coincides on $\mathcal{K}_{H,E}$ with the least upper bound of the sequence $\{f_m\}$ of continuous functions. \smallskip

The  third inequality in (\ref{H-DB-ub}) is obvious.\smallskip

To show that all the bounds in (\ref{H-DB-ub}) are sharp consider the sequence
of states
\begin{equation}\label{sequence}
\rho_{n}=\rho^{\shs\downarrow}_{n}=(1-q_{n})|0\rangle\langle
0|+q_{n}n^{-1}\sum_{k=1}^{n}|k\rangle\langle k|,
\end{equation}
where
$\{q_{n}=(E-E_{0})\left(n^{-1}\sum_{k=1}^{n}E_{k}-E_{0}\right)^{-1}\}$
is a sequence of positive numbers converging to
zero (we assume that $n$ is sufficiently large so that
$q_{n}\leq1$). The sequence $\{\rho_{n}\}$ lies in
$\mathcal{K}_{H,E}$ (since $\Tr H\rho_n=E$) and converges to the pure state
$|0\rangle\langle 0|$. By concavity of the entropy we have
$$
H(\rho_{n})\geq q_{n}\log
n=\frac{(E-E_{0})\log
n}{n^{-1}\sum_{k=1}^{n}E_{k}-E_{0}}\geq\frac{(E-E_{0})\log
n}{E_{n}-E_{0}}.
$$
Since $\,\mathrm{dj}\!\left\{E_H(\rho_n)\right\}=E-E_{0}$, to complete the proof of part A it suffices to note that $\,\limsup_n\log
n(E_{n}-E_{0})^{-1}=g(H)$.\smallskip

B) We have only to prove the existence of a converging sequence $\{\rho_n\}$ such that $\,\mathrm{dj}\!\left\{H(\rho_n)\right\}=+\infty$ in the case $g(H)=+\infty$.
By the remark before the proposition in this case there is a state $\sigma\in\mathcal{K}_{H,E}$ such that $H(\sigma)=+\infty$. The sequence consisting of the states
$\rho_n=n^{-1}\sigma+(1-n^{-1})\rho_0$, where $\rho_0$ is any pure state in $\mathcal{K}_{H,E}$, possesses the required property. $\square$

\section{Estimates for discontinuity  of some information quantities}

In this section we show that discontinuity jumps of many information characteristics of composite quantum states are upper bounded by
 discontinuity jump of one of the marginal entropies (with a corresponding coefficient).

\subsection{Quantum mutual information and conditional entropy}

Quantum mutual information of a state $\,\omega_{AB}\,$ of an
infinite-dimensional bipartite quantum system is defined as follows (cf.\cite{L-mi})
\begin{equation*}
I(A\!:\!B)_{\omega}=H(\omega_{AB}\shs\Vert\shs\omega_{A}\otimes
\omega_{B}).
\end{equation*}
We will use the homogeneous extension of this quantity to positive trace-class operators
\begin{equation*}
I(A\!:\!B)_{\omega}\doteq [\Tr\shs\omega] I(A\!:\!B)_{\frac{\omega}{\Tr\shs\omega}},\quad \omega\in\T_{+}(\H_{AB}).
\end{equation*}
Basic properties of the relative entropy show that $\omega\mapsto
I(A\!:\!B)_{\omega}$ is a lower semicontinuous function on the cone
$\T_{+}(\H_{AB})$ taking values in $[0,+\infty]$. It is easy to show that (cf.\cite{MI-B})
\begin{equation}\label{mi-ub}
I(A\!:\!B)_{\omega}\leq 2\min\left\{H(\omega_A),H(\omega_B)\right\}.
\end{equation}

By the lower semicontinuity of quantum mutual information its discontinuity for a given sequence $\{\omega_{AB}^n\}\subset\T_{+}(\H_{AB})$ converging to an operator $\omega_{AB}^0\in\T_{+}(\H_{AB})$ is characterised by the nonnegative value
$$
\mathrm{dj}\!\left\{I(A\!:\!B)_{\omega^n}\right\}\doteq\limsup_{n\rightarrow+\infty}I(A\!:\!B)_{\omega^n}-I(A\!:\!B)_{\omega^0}
$$
which can be called \emph{mutual information loss} corresponding to this sequence (it is assumed as usual that $\shs\mathrm{dj}\!\left\{I(A\!:\!B)_{\omega^n}\right\}=+\infty\shs$ if $\shs I(A\!:\!B)_{\omega^0}=+\infty$). \smallskip

The following theorem is essentially used below.\smallskip

\begin{theorem}\label{MI-DB} \emph{Let $\{\omega_{AB}^n\}\subset\T_{+}(\H_{AB})$ be a sequence converging to an operator $\omega_{AB}^0$ and $\,\Phi:A\rightarrow C$, $\Psi:B\rightarrow D$ be quantum operations. Then}
\begin{equation*}
\mathrm{dj}\!\left\{I(C\!:\!D)_{\Phi\otimes\Psi(\omega_{AB}^n)}\right\}\leq\mathrm{dj}\!\left\{I(A\!:\!B)_{\omega^n}\right\}\leq 2\min\left\{\mathrm{dj}\!\left\{H(\omega_{A}^n)\right\}, \mathrm{dj}\!\left\{H(\omega_{B}^n)\right\}\right\}.
\end{equation*}
\end{theorem}
\smallskip

Example 1 in \cite{CMI} shows that $\,\mathrm{dj}\!\left\{I(A\!:\!B)_{\omega^n}\right\}$  may vanish despite positivity of $\,\min\left\{\mathrm{dj}\!\left\{H(\omega_{A}^n)\right\}, \mathrm{dj}\!\left\{H(\omega_{B}^n)\right\}\right\}$. On the other hand, by considering sequences of pure states we see that this upper bound is sharp. \smallskip

The first inequality in Theorem \ref{MI-DB} means that \emph{discontinuity jumps of quantum mutual information do not increase under action of local operations}. So, it generalizes Theorem 1B in \cite{CMI} stating  that local continuity of quantum mutual information  is preserved by local operations.\smallskip

\emph{Proof.} To prove the second inequality of the theorem we will use the identity
\begin{equation}\label{sp-ident}
    I(A\!:\!B)_{\omega}+I(A\!:\!C)_{\omega}=2H(\omega_{A})
\end{equation}
valid for any 1-rank operator $\omega\in\T_+(\H_{ABC})$ (with
possible value $+\infty$ in the both sides). If $H(\omega_{A})$,
$H(\omega_{B})$ and $H(\omega_{C})$ are finite then (\ref{sp-ident})
is easily verified by noting that $H(\omega_{A})=H(\omega_{BC})$,
$H(\omega_{B})=H(\omega_{AC})$ and $H(\omega_{C})=H(\omega_{AB})$.
In general case (\ref{sp-ident}) can be proved by approximation (see the proof of Theorem 1 in \cite[the Appendix]{CMI}).\smallskip

It suffices to prove the inequality $\,\mathrm{dj}\!\left\{I(A\!:\!B)_{\omega^n}\right\}\leq 2\mathrm{dj}\!\left\{H(\omega_{A}^n)\right\}$ assuming that $\,H(\omega_{A}^0)<+\infty$.  By Lemma \ref{p-lemma} there is a sequence $\{\tilde{\omega}_{ABC}^n\}$ of
 1-rank operators in $\T_+(\H_{ABC})$ converging to an operator
$\tilde{\omega}^0_{ABC}$ such that $\,\tilde{\omega}^n_{AB}=\omega^n_{AB}$ for
all $n\geq0$. By Lemma \ref{vsl+} identity (\ref{sp-ident}) and the lower semicontinuity of the function $\shs\omega_{ABC}\mapsto I(A\!:\!C)_{\omega}$ imply
the required inequality.\smallskip

To prove the first inequality of the theorem it suffices to show that
\begin{equation}\label{MI-DB-ub}
\mathrm{dj}\!\left\{I(C\!:\!B)_{\Phi\otimes\id_B(\omega_{AB}^n)}\right\}\leq\mathrm{dj}\!\left\{I(A\!:\!B)_{\omega^n}\right\}
\end{equation}
for any quantum operation $\Phi:A\rightarrow C$. We will use the identity (chain rule)
$$
I(A\!:\!B)_{\omega}+I(B\!:\!C|A)_{\omega}=I(AC\!:\!B)_{\omega},
$$
where $I(B\!:\!C|A)_{\omega}$ is the conditional mutual information extended to the cone $\T_{+}(\H_{ABC})$ (see Section 4.3 below).
By Lemma \ref{vsl+} this identity and  the lower semicontinuity of $I(B\!:\!C|A)_{\omega}$ (stated in \cite[Th.2]{CMI})
imply
\begin{equation}\label{MI-DB-ub+}
\mathrm{dj}\!\left\{I(A\!:\!B)_{\omega^n}\right\}\leq\mathrm{dj}\!\left\{I(AC:\!B)_{\omega^n}\right\}
\end{equation}
for any converging sequence  $\{\omega_{ABC}^n\}\subset\T_{+}(\H_{ABC})$. By using the Stinespring representation (\ref{St-rep}) one can show that (\ref{MI-DB-ub+}) implies (\ref{MI-DB-ub}) for any quantum channel $\Phi$.
\smallskip

If $\Phi$ is a trace non-preserving  operation then consider
the channel $\Phi'=\Phi\oplus\Delta$ from $A$ to $C'=C\oplus
C^\mathrm{c}$, where $\Delta(\rho)=[\Tr\rho-\Tr\Phi(\rho)]\sigma$
is a quantum operation from $A$ to $C^\mathrm{c}$ determined by a
fixed state $\sigma\in\S(\H_{C^\mathrm{c}})$. We have
\begin{equation}\label{d-f}
\begin{array}{rl}
I(C'\!:\!B)_{\Psi\otimes\id_B(\omega_{AB})}\!\!&
= I(C\!:\!B)_{\tilde{\omega}}+
H\left(\tilde{\omega}_{B}\shs\Vert\shs\lambda\omega_{B}\right)\\\\&+\,H\left(\Delta\otimes\id_B(\omega_{AB})\shs\Vert\shs\Delta(\omega_{A})\otimes
\omega_{B}\right),
\end{array}
\end{equation}
where $\tilde{\omega}_{CB}=\Phi\otimes\id_B(\omega_{AB})$ and
$\lambda=\Tr\shs\tilde{\omega}_{CB}$ (see the proof of Th.1B in \cite{CMI}).

\smallskip

Since all the summands in the right hand side of
(\ref{d-f}) are lower semicontinuous functions, Lemma \ref{vsl+} implies
$$
\mathrm{dj}\!\left\{I(C\!:\!B)_{\Phi\otimes\id_B(\omega_{AB}^n)}\right\}\leq\mathrm{dj}\!\left\{I(C'\!:\!B)_{\Phi'\otimes\id_B(\omega_{AB}^n)}\right\}
\leq\mathrm{dj}\!\left\{I(A\!:\!B)_{\omega^n}\right\},
$$
where the second inequality holds, since  $\Phi'$ is a channel. $\square$ \medskip

The quantum conditional entropy
\begin{equation}\label{c-e-d}
H(A|B)_{\omega}=H(\omega_{AB})-H(\omega_B)
\end{equation}
can be extended to the convex set
$\S_A\doteq\{\shs\omega_{AB}\,|\,H(\omega_{A})<+\infty\shs\}$ containing
states with $\,H(\omega_{AB})=H(\omega_B)=+\infty\,$ by the formula
\begin{equation}\label{c-e-d+}
H(A|B)_{\omega}=H(\omega_{A})-I(A\!:\!B)_{\omega}
\end{equation}
preserving all basic properties of the conditional entropy \cite{Kuz}. Upper bound (\ref{mi-ub}) shows that
$H(A|B)_{\omega}$ takes values in the interval $\,[-H(\omega_{A}),H(\omega_{A})]$.  \smallskip

The conditional entropy is not upper or lower semicontinuous.\footnote{By Corollary \ref{simple-p-c-3+} in Sec.3.2 the conditional entropy is lower semicontinuous on the set of separable states with finite entropy.} So, its discontinuity jumps
for a given sequence $\{\omega_{AB}^n\}\subset\S_A$ converging to a state $\omega_{AB}^0\in\S_A$ can be characterised by two nonnegative values
$$
\mathrm{dj}^\downarrow\! \left\{H(A|B)_{\omega^n}\right\}\doteq\max\!\left\{\,\limsup_{n\rightarrow+\infty}H(A|B)_{\omega^n}\!-H(A|B)_{\omega^0}\!,\; 0\,\right\}
$$
and
$$
\mathrm{dj}^\uparrow\! \left\{H(A|B)_{\omega^n}\right\}\doteq\max\!\left\{ H(A|B)_{\omega^0}\!-\liminf_{n\rightarrow+\infty}H(A|B)_{\omega^n},\;0\,\right\}
$$
describing respectively the maximal loss and the maximal gain of the conditional entropy corresponding to this sequence.\smallskip

\begin{corollary}\label{CE-DB} \emph{Let $\{\omega_{AB}^n\}$ be a  sequence converging to a state $\omega_{AB}^0$ such that $H(\omega_{A}^n)<+\infty$ for all $\,n\geq0$. Then
\begin{equation}\label{CE-DB-ub}
\mathrm{dj}^\downarrow\! \left\{H(A|B)_{\omega^n}\right\}\leq\min\left\{\mathrm{dj}\!\left\{H(\omega_{A}^n)\right\}, \mathrm{dj}\!\left\{H(\omega_{AB}^n)\right\}\right\},
\end{equation}
\begin{equation*}
\mathrm{dj}^\uparrow\!\left\{H(A|B)_{\omega^n}\right\}\leq \min\left\{2\mathrm{dj}\!\left\{H(\omega_{A}^n)\right\}, \mathrm{dj}\!\left\{H(\omega_{B}^n)\right\}\right\}.
\end{equation*}
If $\{H(\omega_{A}^n)\}$ is a converging sequence then the factor $2$ in the last inequality can be removed.}
\emph{If $\,\omega_{AB}^n, \omega_{AB}^0$ are separable states with finite entropy then $\mathrm{dj}^\uparrow\!\left\{H(A|B)_{\omega^n}\right\}=0$.}
\end{corollary}\medskip

\emph{Proof.} Inequalities (\ref{CE-DB-ub}) and
$\,\mathrm{dj}^\uparrow\!\left\{H(A|B)_{\omega^n}\right\}\leq \mathrm{dj}\!\left\{H(\omega_{B}^n)\right\}\,$
are derived from (\ref{c-e-d}) and (\ref{c-e-d+}) by using Lemma \ref{vsl+} and the lower semicontinuity of $H(\omega_{AB})$, $H(\omega_{B})\shs$ and  $I(A\!:\!B)_{\omega}$.

If $\,\mathrm{dj}^\uparrow\!\left\{H(A|B)_{\omega^n}\right\}>0$ then (\ref{c-e-d+}) implies
$$
\mathrm{dj}^\uparrow\!\left\{H(A|B)_{\omega^n}\right\}\leq\!
\left[\limsup_{n\rightarrow+\infty}I(A\!:\!B)_{\omega^n}-I(A\!:\!B)_{\omega^0}\right]-
\left[\liminf_{n\rightarrow+\infty}H(\omega_{A}^n)-H(\omega_{A}^0)\right]\!.
$$
So, the inequality
$\mathrm{dj}^\uparrow\!\left\{H(A|B)_{\omega^n}\right\}\leq 2\mathrm{dj}\!\left\{H(\omega_{A}^n)\right\}$ follow from Theorem \ref{MI-DB} and the lower semicontinuity of  $H(\omega_{A})$. If $\{H(\omega_{A}^n)\}$ is a converging sequence then $\displaystyle\liminf_{n\rightarrow+\infty}H(\omega_{A}^n)-H(\omega_{A}^0)=\mathrm{dj}\!\left\{H(\omega_{A}^n)\right\}$. \smallskip

The last assertion follows from the lower semicontinuity  of the conditional entropy on the set of separable states with finite entropy (Cor.\ref{simple-p-c-3+} in Sec.3.2). $\square$
\medskip

\subsection{The Holevo quantity of ensemble of quantum states}

The Holevo quantity of an ensemble $\{\pi_i,\rho_i\}$ of quantum states is defined as
$$
\chi(\{\pi_i,\rho_i\})\doteq \sum_i\pi_i H(\rho_i\|\bar{\rho})=H(\bar{\rho})-\sum_i\pi_i H(\rho_i),\quad \bar{\rho}=\sum_i\pi_i\rho_i,
$$
where the second formula is valid if $H(\bar{\rho})<+\infty$. It plays a basic role in analysis of information properties of quantum systems and channels \cite{H-SCI,N&Ch}.

We will say that a sequence $\{\{\pi^n_i,\rho^n_i\}_i\}_n$ of ensembles converges to an ensemble $\{\pi^0_i,\rho^0_i\}$  if
\begin{equation}\label{en-conv}
\lim_{n\rightarrow\infty}\pi^n_i=\pi^0_i\,\textrm{ for all } i\,\textrm{ and }\lim_{n\rightarrow\infty}\rho^n_i=\rho^0_i \,\textrm{ for all } \, i \, \textrm{ s.t. } \pi^0_i\neq0.
\end{equation}
The lower semicontinuity of the relative entropy implies
lower semicontinuity of the Holevo quantity with respect to this convergence. So, its discontinuity  for a sequence $\{\{\pi^n_i,\rho^n_i\}_i\}_n$ converging to an ensemble $\{\pi^0_i,\rho^0_i\}$ is characterised by the nonnegative value
$$
\mathrm{dj}\!\left\{\chi(\{\pi^n_i,\rho^n_i\})\right\}\doteq\limsup_{n\rightarrow+\infty}\chi(\{\pi^n_i,\rho^n_i\})-\chi(\{\pi^0_i,\rho^0_i\})
$$
which can be called \emph{loss of the Holevo quantity} corresponding to this sequence (it is assumed  that $\,\mathrm{dj}\!\left\{\chi(\{\pi^n_i,\rho^n_i\})\right\}=+\infty\,$ if $\,\chi(\{\pi^0_i,\rho^0_i\})=+\infty$).

\smallskip

\begin{property}\label{Chi-DB} \emph{Let $\{\{\pi^n_i,\rho^n_i\}_{i=1}^m\}$ be a sequence of ensembles consisting of $\,m\leq+\infty$ states converging to an ensemble $\{\{\pi^0_i,\rho^0_i\}_{i=1}^m\}$. Then
\begin{equation*}
\mathrm{dj}\!\left\{\chi(\{\pi^n_i,\rho^n_i\}_{i=1}^m)\right\}_n\leq \min\left\{\mathrm{dj}\!\left\{H(\bar{\rho}_n)\right\},2\mathrm{dj}\!\left\{S(\bar{\pi}_n)\right\}\right\},
\end{equation*}
where $\bar{\rho}_n\doteq\sum_{i=1}^m\pi^n_i\rho^n_i$ and $\,\bar{\pi}_n$ is the probability distribution $\{\pi^n_i\}_{i=1}^m$.}\smallskip

\emph{If $\,\lim_{n\rightarrow+\infty}S(\bar{\pi}_n)=S(\bar{\pi}_0)<+\infty$ (in particular, if $\shs m<+\infty$) then}
\begin{equation*}
\mathrm{dj}\!\left\{H\!\left(\sum_{i=1}^m\pi^n_i \rho^n_i\right)\right\}_{\!\!n}=\mathrm{dj}\!\left\{\sum_{i=1}^m\pi^n_i H(\rho^n_i)\right\}_{\!\!n}.
\end{equation*}
\end{property}

\emph{Proof.} To  prove the inequality $\mathrm{dj}\!\left\{\chi(\{\pi^n_i,\rho^n_i\}_{i=1}^m)\right\}_n\leq \mathrm{dj}\!\left\{H(\bar{\rho}_n)\right\}$ we may assume that $H(\bar{\rho}_n)$ is finite for all $n$. So, we have
\begin{equation}\label{tmp}
\chi(\{\pi^n_i,\rho^n_i\}_{i=1}^m)+\sum_{i=1}^m\pi^n_i H(\rho^n_i)=H(\bar{\rho}_n)\quad \forall n.
\end{equation}
Thus, the required inequality follows from Lemma \ref{vsl+} and the lower semicontinuity of the second term in (\ref{tmp}) with respect to the convergence (\ref{en-conv}).

\smallskip

To prove the inequality $\,\mathrm{dj}\!\left\{\chi(\{\pi^n_i,\rho^n_i\}_{i=1}^m)\right\}_n\leq 2\mathrm{dj}\!\left\{S(\bar{\pi}_n)\right\}\,$ assume that  $\H_A=\H$ and $\,\H_B=\mathbb{C}^m$. It is easy to see that $\,\chi(\{\pi^n_i,\rho^n_i\}_{i=1}^m)=I(A\!:\!B)_{\omega^n}$ for each $n\geq0$, where
\begin{equation}\label{qc-states}
\omega_{AB}^n=\sum_{i=1}^m\pi^n_i\rho^n_i\otimes |i\rangle\langle i|
\end{equation}
is a state in $\S(\H_{AB})$ determined by a basis $\{|i\rangle\}$ in $\H_B$. Since $\,H(\omega_{B}^n)=S(\bar{\pi}_n)\,$ for all $n$, to obtain the required inequality from Theorem \ref{MI-DB} it suffices to show convergence of the sequence $\{\omega_{AB}^n\}$ to the state $\omega_{AB}^0$. This can be done by noting that (\ref{en-conv}) implies convergence of the sequence $\{\omega_{AB}^n\}$ to the state $\omega_{AB}^0$ in the weak operator topology and by using the result from \cite{D-A}. \smallskip

The second assertion of the proposition follows from the first one.  $\square$ \smallskip

Proposition \ref{Chi-DB}B implies the following strengthened version of Corollary \ref{appr-DB+}.\smallskip

\begin{corollary}\label{Chi-DB+}
\emph{Let $\{\rho^1_n\}_n,...,\{\rho^m_n\}_n$ be sequences of operators in  $\T_+(\H)$ converging to operators
$\rho^1_0,...,\rho^m_0$, where $m\leq+\infty$. The equality
\begin{equation}\label{Chi-DB-ub++}
\mathrm{dj}\!\left\{H\!\left(\sum_{k=1}^m\rho^k_n\right)\right\}_{\!\!n}=\,\mathrm{dj}\!\left\{\sum_{k=1}^m H(\rho^k_n)\right\}_{\!\!n}
\end{equation}
holds if
$$
\!\!\lim_{n\rightarrow\infty}\sum_{k=1}^m\Tr \rho_n^k=\sum_{k=1}^m\Tr\rho_0^k<\infty\;\;\textrm{and}\;\; \lim_{n\rightarrow\infty}\!S(\{\Tr\rho^k_n\}_{k=1}^m)=S(\{\Tr\rho^k_0\}_{k=1}^m)<\infty,
$$
where $\,S(\{x_k\})\doteq\sum_k\eta(x_k)-\eta(\sum_kx_k)\,$ is the homogeneous extension of the Shannon entropy to the positive cone of $\,\ell_1$. In particular, relation (\ref{Chi-DB-ub++}) holds if $\,m<+\infty$.}
\end{corollary}

\subsection{Conditional mutual information}

The conditional mutual information of a state $\omega_{ABC}$ of a
tripartite finite-dimensional system  is defined as follows
\begin{equation}\label{cmi-d}
    I(A\!:\!C|B)_{\omega}\doteq
    H(\omega_{AB})+H(\omega_{BC})-H(\omega_{ABC})-H(\omega_{B}).
\end{equation}
This quantity plays important role in quantum
information theory \cite{D&J, Brandao}, its nonnegativity is a basic result well known as \emph{strong subadditivity
of von Neumann entropy} \cite{Simon}.

In infinite dimensions formula (\ref{cmi-d}) may contain the uncertainty
$"\infty-\infty"$. Nevertheless the
conditional mutual information can be defined for any state
$\omega_{ABC}$ by one of the equivalent expressions
\begin{equation}\label{cmi-e+}
\!I(A\!:\!C|B)_{\omega}=\sup_{P_A}\left[\shs I(A\!:\!BC)_{Q_A\omega
Q_A}-I(A\!:\!B)_{Q_A\omega Q_A}\shs\right],\; Q_A=P_A\otimes I_{BC},\!
\end{equation}
\begin{equation}\label{cmi-e++}
\!I(A\!:\!C|B)_{\omega}=\sup_{P_C}\left[\shs I(AB\!:\!C)_{Q_C\omega
Q_C}-I(B\!:\!C)_{Q_C\omega Q_C}\shs\right],\; Q_C=P_C\otimes I_{AB},\!
\end{equation}
where the suprema are over all finite rank projectors
$P_A\in\B(\H_A)$ and\break $P_C\in\B(\H_C)$ correspondingly \cite{CMI}.\smallskip

It is shown in \cite[Th.2]{CMI} that expressions (\ref{cmi-e+}) and
(\ref{cmi-e++}) define a lower semicontinuous function on the set
$\S(\H_{ABC})$ possessing all basic properties of conditional mutual
information valid in finite dimensions. If one of the marginal
entropies $H(\omega_A)$, $H(\omega_C)$ and $H(\omega_B)$ is finite
then the above extension is given respectively by the explicit formula\footnote{The correctness of these formulas follows from upper bound (\ref{mi-ub}).}
\begin{equation}\label{cmi-d+}
I(A\!:\!C|B)_{\omega}=I(A\!:\!BC)_{\omega}-I(A\!:\!B)_{\omega},
\end{equation}
\begin{equation}\label{cmi-d++}
I(A\!:\!C|B)_{\omega}=I(AB\!:\!C)_{\omega}-I(B\!:\!C)_{\omega}
\end{equation}
and
\begin{equation}\label{cmi-d+++}
I(A\!:\!C|B)_{\omega}=I(A\!:\!C)_{\omega}-I(A\!:\!B)_{\omega}-I(B\!:\!C)_{\omega}+I(AC\!:\!B)_{\omega}.
\end{equation}

Since $\,\omega_{ABC}\mapsto
I(A\!:\!C|B)_{\omega}\,$ is a lower semicontinuous function, its discontinuity for a given sequence $\{\omega_{ABC}^n\}$ of states converging to a state $\omega_{ABC}^0$ with finite $\,I(A\!:\!C|B)_{\omega^0}$  is characterised by the nonnegative value
$$
\mathrm{dj}\!\left\{I(A\!:\!C|B)_{\omega^n}\right\}\doteq\limsup_{n\rightarrow+\infty}I(A\!:\!C|B)_{\omega^n}-I(A\!:\!C|B)_{\omega^0}.
$$

\begin{property}\label{CMI-DB} \emph{For an arbitrary sequence $\{\omega_{ABC}^n\}$ of states converging to a state $\omega_{ABC}^0$ the following inequalities hold}
\begin{equation*}
\begin{array}{l}
\!\mathrm{dj}\!\left\{I(A\!:\!C|B)_{\omega^n}\right\}\leq 2\min\left\{\mathrm{dj}\!\left\{H(\omega_{A}^n)\right\}\!, \mathrm{dj}\!\left\{H(\omega_{C}^n)\right\}\!, \mathrm{dj}\!\left\{H(\omega_{AB}^n)\right\}\!, \mathrm{dj}\!\left\{H(\omega_{BC}^n)\right\}\right\}\!,\\\\
\!\mathrm{dj}\!\left\{I(A\!:\!C|B)_{\omega^n}\right\}\leq \mathrm{dj}\!\left\{I(A\!:\!C)_{\omega^n}\right\}+2\min\left\{\mathrm{dj}\!\left\{H(\omega_{B}^n)\right\},\mathrm{dj}\!\left\{H(\omega_{ABC}^n)\right\}\right\}.
\end{array}
\end{equation*}
\end{property}

\smallskip

\emph{Proof.} The first inequality is derived from representations (\ref{cmi-d+}) and (\ref{cmi-d++}) by using  Lemma  \ref{vsl+}, the lower semicontinuity of the quantum mutual information and Theorem \ref{MI-DB}. The inequality
$$
\mathrm{dj}\!\left\{I(A\!:\!C|B)_{\omega^n}\right\}\leq \mathrm{dj}\!\left\{I(A\!:\!C)_{\omega^n}\right\}+2\mathrm{dj}\!\left\{H(\omega_{B}^n)\right\}
$$
 is derived by the same way from representation (\ref{cmi-d+++}).

To prove the inequality $\,\mathrm{dj}\!\left\{I(A\!:\!C|B)_{\omega^n}\right\}\leq \mathrm{dj}\!\left\{I(A\!:\!C)_{\omega^n}\right\}+2\mathrm{dj}\!\left\{H(\omega_{ABC}^n)\right\}$
note that Lemma \ref{p-lemma} implies existence of a sequence $\{\widetilde{\omega}_{ABCD}^n\}$ of pure states converging to a state  $\shs\widetilde{\omega}_{ABCD}^0$ such that
$\shs\widetilde{\omega}_{ABC}^n=\omega_{ABC}^n$ for all $\shs n\geq0$. We have  (cf.\cite{D&J})
$$
I(A\!:\!C|B)_{\omega^n}=I(A\!:\!C|D)_{\tilde{\omega}^n}\quad\textrm{and}\quad H(\omega_{ABC}^n)=H(\tilde{\omega}_D^n)\; \textrm{ for all }\; n\geq0.
$$
So, the required inequality is derived from representation (\ref{cmi-d+++})
applied to\break $I(A\!:\!C|D)$ by using   Lemma  \ref{vsl+},  the lower semicontinuity of the quantum mutual information and Theorem \ref{MI-DB}. $\square$

\subsection{Several entanglement measures}

In this subsection we will obtain estimates for discontinuity jumps of the infinite-dimensional versions of
squashed entanglement, c-squashed entanglement, entanglement of formation and of their regularizations. \smallskip

The \emph{squashed entanglement} $E_{sq}$ of a state $\omega_{AB}$ of a finite
dimensional bipartite system  is defined as follows
\begin{equation}\label{se-def}
  E_{sq}(\omega_{AB})=\textstyle\frac{1}{2}\displaystyle\inf_{\omega_{ABE}}I(A\!:\!B|E),
\end{equation}
where $I(A\!:\!B|E)_{\omega}$ is the conditional mutual information defined by (\ref{cmi-d}) and the infimum is over all extensions $\,\omega_{ABE}\,$ of the state
$\,\omega_{AB}$ \cite{C&W,Tucci}. The squashed entanglement is an unique known entanglement measure possessing
all  basic properties of an entanglement
measure including additivity and monogamy \cite{C&W,SE-F,K&W}.

Possible generalizations of squashed entanglement to states of infinite-dimensional bipartite system are considered in \cite{SE}, where it is shown
that it can be unambiguously  defined on the set
\begin{equation}\label{star}
\S_\mathrm{*}\doteq\{\shs\omega_{AB}\,|\,\min\{H(\omega_{A}),H(\omega_{B}),H(\omega_{AB})\}<+\infty\shs\}
\end{equation}
by the same formula (\ref{se-def}), where $I(A\!:\!B|E)_{\omega}$ is the extended conditional mutual information described in the previous subsection, as a lower semicontinuous entanglement measure
possessing all basic properties of the
squashed entanglement valid in finite dimensions.\smallskip

\begin{remark}\label{un-ext}
It is shown in \cite{SE} that any continuous finite-dimensional entanglement measure has an unique lower semicontinuous extension to the set of all infinite-dimensional bipartite states, but it is not clear how to prove coincidence of this "universal" extension with the quantity obtained by direct translation of the finite-dimensional definition. The above set $\S_\mathrm{*}$ is the maximal set of states on which such coincidence is proved for the squashed entanglement (as well as for the c-squashed entanglement and for the entanglement of formation considered below). $\square$
\end{remark} \smallskip

The \emph{c-squashed entanglement} $E_{csq}$ of a state $\omega_{AB}$ of a finite-dimensional bipartite system  is defined by the formula
\begin{equation}\label{csq-def}
  E_{csq}(\omega_{AB})=\textstyle\frac{1}{2}\displaystyle\inf_{\omega_{ABE}\in\S_\mathrm{c}}I(A\!:\!B|E),
\end{equation}
where $\S_\mathrm{c}$ is the set of all extensions of $\omega_{AB}$ having the form
\begin{equation}\label{form}
\omega_{ABE}=\sum_i \pi_i\shs\omega^i_{AB}\otimes|i\rangle\langle i|_E.
\end{equation}
This means  that
\begin{equation}\label{csq-def+}
  E_{csq}(\omega_{AB})=\inf_{\sum_i\pi_i\omega_{AB}^i=\omega_{AB}}\sum_i \pi_i I(A\!:\!B)_{\omega^i},
\end{equation}
where the infimum is over all ensembles $\{\pi_i, \omega_{AB}^i\}$ of states with the average state $\omega_{AB}$ \cite{Nagel,4H}, i.e.
$E_{csq}$ is the convex hull (mixed convex roof) of the quantum mutual information.

An universal infinite-dimensional extension (mentioned in Remark \ref{un-ext}) of the c-squashed entanglement is given by the formula
\begin{equation}\label{csq-def++}
  E_{csq}(\omega_{AB})=\inf_{b(\mu)=\omega_{AB}}\int\! I(A\!:\!B)_{\omega}\mu(d\omega),
\end{equation}
where the infimum is over all Borel probability measures on the set $\S(\H_{AB})$ with the barycenter $\omega_{AB}$.\footnote{The integral is well defined for any such $\mu$ due to the lower semicontinuity of $I(A\!:\!B)_{\omega}$.} Indeed, Proposition 1 in \cite{EM} and Corollary  1 in \cite{EM} imply lower semicontinuity of the right hand side of (\ref{csq-def++})  and its coincidence with
the right hand side of (\ref{csq-def+}) for any state $\omega_{AB}$ with finite rank marginals.

By using Corollary 6  in \cite{EM} and upper bound (\ref{mi-ub}) one can show that formulas (\ref{csq-def+}) and (\ref{csq-def++}) coincide for any
state $\omega_{AB}$ in the set $\S_\mathrm{*}$ defined in (\ref{star}). So, representation (\ref{csq-def}) remains valid for any $\omega_{AB}\in\S_\mathrm{*}$ and hence $E_{csq}$ is not less than $E_{sq}$ on $\S_\mathrm{*}$.  Global coincidence of (\ref{csq-def+}) and (\ref{csq-def++})
is an open question. By Proposition 1 in \cite{SE} it is  equivalent  to global lower semicontinuity of the right hand side of (\ref{csq-def+}).\smallskip

The \emph{entanglement of formation}  of a state $\omega_{AB}$ of a finite
dimensional bipartite system is defined as follows
\begin{equation}\label{ef-def}
  E_{F}(\omega_{AB})=\inf_{\sum_i\pi_i\omega_{AB}^i=\omega_{AB}}\sum_i \pi_iH(\omega^i_{A}),
\end{equation}
where the infimum is over all ensembles $\{\pi_i, \omega_{AB}^i\}$ of pure states with the average state $\omega_{AB}$ \cite{Bennett}.
The entanglement of formation  is one of the most important entanglement measures -- it is the maximal convex continuous function coinciding with the marginal entropy of a state on the set of pure bipartite states \cite{4H,P&V}.

An universal infinite-dimensional extension (mentioned in Remark \ref{un-ext}) of the entanglement of formation  is given by the formula
\begin{equation}\label{ef-def+}
E_F(\omega_{AB})=\!\inf_{b(\mu)=\omega_{AB}}\int\! H(\omega_A)\mu(d\omega),
\end{equation}
where the infimum is over all Borel probability measures on the set $\mathrm{ext}\shs\S(\H_{AB})$ of pure states with the barycenter $\omega_{AB}$ (see the end of Sect.3 in \cite{SE}). Similar to the case of $E_{csq}$  formulas (\ref{ef-def}) and (\ref{ef-def+}) coincide for any
state $\omega_{AB}$ in the set $\S_\mathrm{*}$ defined in (\ref{star}). Global coincidence of (\ref{ef-def}) and (\ref{ef-def+})
is a conjecture equivalent to global lower semicontinuity of the right hand side of (\ref{ef-def}).

Since the right hand side of (\ref{ef-def}) can be written as the right hand side of (\ref{csq-def}) with the set $\S_\mathrm{c}$ replaced by its subset $\S^\mathrm{p}_\mathrm{c}$ consisting of all states (\ref{form}) such that $\rank\shs\omega^i_{AB}=1$ for all $i$ (cf.\cite{C&W}), we have
$$
E_{sq}(\omega_{AB})\leq E_{csq}(\omega_{AB})\leq E_{F}(\omega_{AB})
$$
for any state $\omega_{AB}$ in $\S_\mathrm{*}$. Examples showing that $"<"$ may hold in the above inequalities are presented in \cite{Brandao, C&W}.  Note also that $E_{sq}$, $E_{csq}$ and $E_{F}$ have the common continuity bound under the energy constraint on one subsystem provided the corresponding Hamiltonian satisfies condition (\ref{Gh}). For $E_{sq}$ and $E_{F}$ this continuity bound is obtained in \cite{SE}, the case of $E_{csq}$ is considered similarly. This bound implies the asymptotic continuity of all these entanglement measures under the energy constraint on one subsystem.

\medskip

In contrast to the squashed entanglement $E_{sq}$, the measures $E_{csq}$ and $E_{F}$ are nonadditive. To obtain additive measures consider the regularizations
$$
E^{\infty}_{csq}(\omega_{AB})\doteq \lim_{k\rightarrow \infty}k^{-1}E_{csq}(\omega^{\otimes k}_{AB}),\qquad
E^{\infty}_{F}(\omega_{AB})\doteq \lim_{k\rightarrow \infty}k^{-1}E_{F}(\omega^{\otimes k}_{AB}).
$$
In finite dimensions  $E^{\infty}_{F}(\omega_{AB})$ coincides with the entanglement cost $E_{C}(\omega_{AB})$  -- an operationally defined entanglement measure \cite{Hayden&Co}.\smallskip

Since $E_{sq}$, $E_{csq}$ and $E_{F}$ are lower semicontinuous functions on the set
$\S_\mathrm{*}$,  discontinuity jumps of these functions for a given sequence $\{\omega_{AB}^n\}\subset\S_\mathrm{*}$ converging to a state $\omega_{AB}^0\in\S_\mathrm{*}$ are characterised by the nonnegative values
\begin{equation}\label{E-fun}
\mathrm{dj}\!\left\{E(\omega_{AB}^n)\right\}\doteq\limsup_{n\rightarrow+\infty}E(\omega_{AB}^n)-E(\omega_{AB}^0),\qquad E=E_{sq},E_{csq},E_{F},
\end{equation}
where it is assumed as usual that $\,\mathrm{dj}\!\left\{E(\omega_{AB}^n)\right\}=+\infty\,$ if $\,E(\omega_{AB}^0)=+\infty$. \smallskip

Lower semicontinuity of the functions $E^{\infty}_{csq}$ and $E^{\infty}_{F}$ on the set
$\S_\mathrm{*}$ is conjectured but not proved.\footnote{Recently Winter proved the continuity of $E^{\infty}_{F}$ of the set $\S(\H_{AB})$ if one of the systems $A$ and $B$ is finite-dimensional \cite{Winter}. So, to  prove the lower semicontinuity of $E^{\infty}_{F}$ on the set $\S_\mathrm{*}$ it suffices to show that  $\sup_n \lambda_nE^{\infty}_{F}(\omega_{AB}^n)=E^{\infty}_{F}(\omega_{AB})$, where $\lambda_n=\Tr P^n_A\otimes I_B\shs\omega_{AB}$, $\omega_{AB}^n=\lambda^{-1}_nP^n_A\otimes I_B\shs\omega_{AB}P^n_A\otimes I_B$, for all $\omega_{AB}\in\S_\mathrm{*}$ and some sequence $\{P^n_A\}$ of finite rank projectors strongly converging to the identity operator $I_A$.} Nevertheless, we can consider the values   $\mathrm{dj}\!\left\{E^{\infty}_{csq}(\omega_{AB}^n)\right\}$ and $\mathrm{dj}\!\left\{E^{\infty}_{F}(\omega_{AB}^n)\right\}$ defined by formula (\ref{E-fun}) with $E=E^{\infty}_{csq},E^{\infty}_{F}$ characterizing maximal loss of these functions for a given converging sequence $\{\omega_{AB}^n\}\subset\S_\mathrm{*}$.

\smallskip

Proposition 2 in \cite{SE} and Proposition 8 in \cite{EM} show that the functions  $E_{sq}$ and $E_{F}$ are continuous on any subset of
$\S(\H_{AB})$ on which one of the marginal entropies $H(\omega_{A})$ and $H(\omega_{B})$  is continuous.  The same condition is valid for
$E_{csq}$. These observations are generalized in the following\smallskip

\begin{property}\label{E-DB} \emph{For an arbitrary sequence $\{\omega_{AB}^n\}\subset\S_\mathrm{*}$ converging to a state $\omega_{AB}^0\in\S_\mathrm{*}$ the following inequalities hold}
\begin{equation*}
\begin{array}{l}
\mathrm{dj}\!\left\{E(\omega_{AB}^n)\right\}\leq \min\left\{\mathrm{dj}\!\left\{H(\omega_{A}^n)\right\}, \mathrm{dj}\!\left\{H(\omega_{B}^n)\right\}\right\},\quad E=E_{sq},E_{csq},E^{\infty}_{csq},E_{F},E^{\infty}_{F}\\\\
\mathrm{dj}\!\left\{E(\omega_{AB}^n)\right\}\leq \textstyle\frac{1}{2}\mathrm{dj}\!\left\{I(A\!:\!B)_{\omega^n})\right\},\qquad\qquad\quad\quad\quad\, E=E_{sq},E_{csq},E^{\infty}_{csq}.
\end{array}
\end{equation*}
\end{property} \smallskip

\emph{Proof.}  The first inequality for $E=E_{F}$ follows from Remark \ref{IQ-DB-r} in Section 5.2 below, since $E_{F}(\omega_{AB})=\overline{\mathrm{co}}H_{\Phi}(\omega_{AB})$, where $\Phi(\omega_{AB})=\omega_{A}$,  for any state $\omega_{AB}\in\S_\mathrm{*}$ \cite{EM}. \smallskip

The first inequality for $E=E_{csq}$, $E_{sq}$ follows from the second one and Theorem \ref{MI-DB}.

To prove the second inequality  for $\shs E=E_{sq}$ assume that $\shs I(A\!:\!B)_{\omega^n}<+\infty$ for all $n$ and  consider
the nonincreasing  sequence of functions
\begin{equation*}
  E^k_{sq}(\omega_{AB})=\textstyle\frac{1}{2}\displaystyle\inf_{\omega_{ABE}}I(A\!:\!B|E)_{\omega},\quad \dim\H_E\leq k
\end{equation*}
pointwise converging to the function $E_{sq}$ on $\S_*$ \cite[Lemma 4]{SE}. By Lemma \ref{vsl} in Section 2 it suffices to show that
$\,\mathrm{dj}\!\left\{E^k_{sq}(\omega_{AB}^n)\right\} \leq \frac{1}{2}\mathrm{dj}\!\left\{I(A\!:\!B)_{\omega^n}\right\}$ for all $k$.

Let $\H^k_E$ be a $k$-dimensional Hilbert space and $\,\widetilde{\omega}^0_{ABE}\in\S(\H_{AB}\otimes\H^k_E)\,$ be an extension of the state
$\omega^0_{AB}$ such that $\,E^k_{sq}(\omega_{AB}^0)\geq \textstyle\frac{1}{2}I(A\!:\!B|E)_{\widetilde{\omega}^0}-\varepsilon$. By using Lemma \ref{p-lemma} it is easy to show existence of a sequence
$\{\tilde{\omega}^n_{ABE}\}$ in $\S(\H_{AB}\otimes\H^k_E)$  converging to the state
$\widetilde{\omega}^0_{ABE}$ such that $\widetilde{\omega}^n_{AB}=\omega^n_{AB}$ for all $n$.

Since $E^k_{sq}(\omega_{AB}^n)\leq \textstyle\frac{1}{2}I(A\!:\!B|E)_{\widetilde{\omega}^n}$ for all $n$, the second inequality in Proposition \ref{CMI-DB} implies
$$
\mathrm{dj}\!\left\{E^k_{sq}(\omega_{AB}^n)\right\}\leq\textstyle\frac{1}{2}\mathrm{dj}\!\left\{I(A\!:\!B|E)_{\widetilde{\omega}^n}\right\}
+\varepsilon\leq\textstyle\frac{1}{2}\mathrm{dj}\!\left\{I(A\!:\!B)_{\omega^n}\right\}+\varepsilon.
$$

The case $E=E_{csq}$ is considered similarly. By using Corollary 6  in \cite{EM} and upper bound (\ref{mi-ub}) one can show that the sequence of functions
$$
E_{csq}^k(\omega_{AB})\doteq\textstyle\frac{1}{2}\displaystyle\inf_{\omega_{ABE}\in \S^k_{\mathrm{c}}}\shs
I(A\!:\!B|E)_{\omega}=\inf_{\sum_{i=1}^k\pi_i\omega_{AB}^i=\omega_{AB}}\sum_{i=1}^k \pi_i I(A\!:\!B)_{\omega^i}
$$
where $\S^k_{\mathrm{c}}$  is the subset of $\S_{\mathrm{c}}$ consisting of states (\ref{form}) with number of summands $\leq k$, pointwise converges to the function $E_{csq}$ on $\S_*$. It suffices only to show existence of a sequence $\{\tilde{\omega}^n_{ABE}\}\subset\S^k_{\mathrm{c}}$ converging to a given state
$\widetilde{\omega}^0_{ABE}\in\S^k_{\mathrm{c}}$ such that $\widetilde{\omega}^n_{AB}=\omega^n_{AB}$ for all $n$.
But this follows from stability of the set $\S(\H_{AB})$ \cite{SSP}, since it implies that for an ensemble $\{\pi^0_i, \rho^0_i\}_{i=1}^k$ with the average state $\rho_0$ and a sequence
$\{\rho_n\}$ converging to the state $\rho_0$ there exits a sequence $\{\{\pi^n_i, \rho^n_i\}_{i=1}^k\}_n$ of ensembles such that $\sum_{i=1}^k\pi^n_i\rho^n_i=\rho_n$ converging to the ensemble $\{\pi^0_i, \rho^0_i\}_{i=1}^k$ in the sense of (\ref{en-conv}).

Consider the cases $E=E^{\infty}_{csq},E^{\infty}_{F}$. Since the functions $E_{csq}$ and $E_{F}$ are subadditive for tensor product states, the
functions $E^{\infty}_{csq}$ and $E^{\infty}_{F}$ are pointwise limits of the non-increasing sequences of functions
$$
E^{k}_{csq}(\omega_{AB})\doteq k^{-1}E_{csq}(\omega^{\otimes k}_{AB}),\qquad
E^{k}_{F}(\omega_{AB})\doteq k^{-1}E_{F}(\omega^{\otimes k}_{AB}).
$$
By Lemma \ref{vsl} in Section 2 to prove the required estimates for $\,\mathrm{dj}\!\left\{E^{\infty}_{csq}(\omega_{AB}^n)\right\}$ and $\mathrm{dj}\!\left\{E^{\infty}_{F}(\omega_{AB}^n)\right\}$ it suffices to show that
$$
\mathrm{dj}\!\left\{E(\omega_{AB}^n)\right\}\leq \min\left\{\mathrm{dj}\!\left\{H(\omega_{A}^n)\right\}, \mathrm{dj}\!\left\{H(\omega_{B}^n)\right\}\right\},\quad E=E^k_{csq},E^k_{F}
$$
and that
$$
\mathrm{dj}\!\left\{E^k_{csq}(\omega_{AB}^n)\right\}\leq \textstyle\frac{1}{2}\mathrm{dj}\!\left\{I(A\!:\!B)_{\omega^n}\right\}
$$
for all $k$. But these relations follow from the same relations with $k=1$ proved before due to the additivity of the von Neumann entropy and of the quantum mutual information. $\square$

\subsection{The Henderson-Vedral measure of classical correlations and quantum discord}

To describe classical component of correlation  of a state $\omega_{AB}$ of a finite
dimensional bipartite system Henderson and Vedral intoduced in \cite{H&V}  the notion of a measure of classical correlations
(as a function satisfying several basic requirements). They also proposed an example of such measure defined as follows
\begin{equation}\label{m-c-c}
C_B(\omega_{AB})=\sup_{\{M_i\}}\left[ H(\omega_{A})-\sum_i\pi_iH(\omega^i_{A})\right],
\end{equation}
where the supremum  is taken over all  measurements (POVM)
$\{M_i\}$ applied to the  system $B$, $\pi_i=\Tr[(I_A\otimes M_i)\omega_{AB}]$
is the probability of the outcome $i$, $\omega^i_{A}=\pi^{-1}_i\Tr_B[(I_A\otimes M_i)\omega_{AB}]$ is the posteriori state of the system $A$ corresponding to  the outcome $i$.

The function $C_B(\omega_{AB})$ is nonnegative, invariant under local unitary trasformations
and  non-increasing under local operations. It coincides with the von Neumann entropy on the set of pure states and with the quantum mutual information on the set of classical-quantum states having form (\ref{qc-states}) \cite{H&V,MI-B,Xi+}.

\smallskip

\begin{property}\label{I-lsc}
\emph{The function $C_B(\omega_{AB})$ is lower semicontinuous on the set $\S(\H_{AB})$ and
\begin{equation}\label{C-DB}
\mathrm{dj}\!\left\{C_B(\omega^n_{AB})\right\}\doteq\limsup_{n\rightarrow+\infty}C_B(\omega^n_{AB})-C_B(\omega^0_{AB})
\leq\mathrm{dj}\!\left\{H(\omega^n_{A})\right\}
\end{equation}
for any sequence $\{\omega^n_{AB}\}\subset\S(\H_{AB})$ converging to a state $\omega^0_{AB}$.} \smallskip

\emph{In particular, local continuity of $H(\omega_{A})$ implies local continuity of $\shs C_B(\omega_{AB})$.}
\end{property}
\medskip

\emph{Proof.} Since for any given measurement $\{M_i\}$ the value in the square bracket in (\ref{m-c-c}) is a lower semicontinuous function of $\omega_{AB}$, the lower semicontinuity of $C_B(\omega_{AB})$ follows from its definition.

To prove (\ref{C-DB}) we will use the Koashi-Winter relation
\begin{equation}\label{K-W}
  C_B(\omega_{AB})+E^d_F(\omega_{AC})=H(\omega_A)
\end{equation}
valid for any pure state $\omega_{ABC}$ \cite{K&W}, where $E^d_F$ is a discrete version of the entanglement of formation defined by formula (\ref{ef-def}).\footnote{A generalizations of the proof of (\ref{K-W})  to infinite dimensions is straightforward.}

We may assume that $H(\omega^n_{A})$ is finite for all $n$.  By Lemma \ref{p-lemma} there is a sequence $\{\tilde{\omega}_{ABC}^n\}$ of
pure states converging to a state $\tilde{\omega}^0_{ABC}$ such that $\,\tilde{\omega}^n_{AB}=\omega^n_{AB}$ for
all $n\geq0$. Proposition 8 in \cite{EM} and the assumed finiteness of $H(\omega^n_{A})$ show that
$$
\liminf_{n\rightarrow+\infty}E^d_F(\omega^n_{AC})\geq E^d_F(\omega^0_{AC}).
$$
So, Lemma \ref{vsl+}  and identity (\ref{K-W}) imply (\ref{C-DB}). $\square$\smallskip

The \emph{quantum discord}  is the difference between the quantum mutual information and the above measure of classical correlations:
\begin{equation}\label{q-d}
D_B(\omega_{AB})\doteq I(A\!:\!B)_{\omega}-C_B(\omega_{AB}).
\end{equation}
It is proposed in \cite{O&Z} as quantity describing quantum component of correlations of a state $\omega_{AB}$ (see \cite{MI-B,Xi+,Str} and the references therein).

The quantum discord is not upper or lower semicontinuous. So, its discontinuity
for a given sequence $\{\omega_{AB}^n\}$ converging to a state $\omega_{AB}^0$ such that $\,I(A\!:\!B)_{\omega_n}<+\infty\,$  can be characterised by two nonnegative values
$$
\mathrm{dj}^\downarrow\! \left\{D_B(\omega_{AB}^n)\right\}\doteq\max\!\left\{\,\limsup_{n\rightarrow+\infty}D_B(\omega_{AB}^n)\!-D_B(\omega_{AB}^0)\!,\; 0\,\right\}
$$
and
$$
\mathrm{dj}^\uparrow\! \left\{D_B(\omega_{AB}^n)\right\}\doteq\max\!\left\{D_B(\omega_{AB}^0)\!-\liminf_{n\rightarrow+\infty}D_B(\omega_{AB}^n),\;0\,\right\}
$$
describing respectively the maximal loss and the maximal gain of the quantum discord corresponding to this sequence.\smallskip

\begin{corollary}\label{QD-DB} \emph{Let $\{\omega_{AB}^n\}$ be a  sequence converging to a state $\omega_{AB}^0$ such that $\,I(A\!:\!B)_{\omega_n}<+\infty$ for all $\,n\geq0$. Then}
\begin{equation}\label{QD-DB-ub}
\begin{array}{l}
\mathrm{dj}^\downarrow\! \left\{D_B(\omega_{AB}^n)\right\}\leq \min\left\{2\mathrm{dj}\!\left\{H(\omega_{A}^n)\right\},\mathrm{dj}\!\left\{H(\omega_{B}^n)\right\}\right\}\\\\
\mathrm{dj}^\uparrow\!\left\{D_B(\omega_{AB}^n)\right\}\leq \min\left\{\mathrm{dj}\!\left\{H(\omega_{A}^n)\right\},\mathrm{dj}\!\left\{H(\omega_{AB}^n)\right\}\right\}.
\end{array}
\end{equation}
\emph{In particular, local continuity of $H(\omega_{A})$ implies local continuity of $\shs D_B(\omega_{AB})$.}
\end{corollary}\medskip

\emph{Proof.} All the upper bounds in (\ref{QD-DB-ub}) are proved by applying Lemma \ref{vsl+}
to relation (\ref{q-d}) and to the following modification of Koashi-Winter relation
\begin{equation*}
  D_B(\omega_{AB})+C_B(\omega_{BC})=H(\omega_B)
\end{equation*}
valid for any pure state $\omega_{ABC}$ \cite{Xi+}, and by using Theorem \ref{MI-DB}, Proposition \ref{I-lsc}, Lemma \ref{p-lemma} and the equality $H(\omega_{AB})=H(\omega_{C})$ for a pure state $\omega_{ABC}$. $\square$

\section{Entropic characteristics of quantum channels and operations}

\subsection{Output entropy of quantum  operations}

The output entropy $H_{\Phi}(\rho)\doteq H(\Phi(\rho))$ of a quantum operation $\Phi:A\rightarrow B$ is
a lower semicontinuous function on the set $\S(\H_A)$ of input states. So, its discontinuity  for a given sequence
$\{\rho_n\}\subset\S(\H_A)$ converging to a state $\rho_0\in\S(\H_A)$ is characterised by the nonnegative value
$$
\mathrm{dj}\!\left\{H_{\Phi}(\rho_n)\right\}\doteq\limsup_{n\rightarrow+\infty}H_{\Phi}(\rho_n)-H_{\Phi}(\rho_0),
$$
which can be called \emph{the output entropy loss} of the operation $\Phi$ corresponding to this sequence (it is assumed that $\,\mathrm{dj}\!\left\{H_{\Phi}(\rho_n)\right\}=+\infty\,$ if $\,H_{\Phi}(\rho_0)=+\infty$). \smallskip

In general, finiteness and local continuity of the von Neumann entropy are not preserved by quantum operations, which means that we can not write general bound for $\,\mathrm{dj}\!\left\{H_{\Phi}(\rho_n)\right\}\,$ in terms of $\,\mathrm{dj}\!\left\{H(\rho_n)\right\}$. The first part of the following theorem characterizes a class of quantum operations for which such bound exists.

\begin{theorem}\label{DB-OE}
A) \textit{Let $\,\Phi:A\rightarrow B$ be a quantum operation. The following
properties are equivalent:}
\begin{enumerate}[(i)]
    \item \emph{there is $\shs C>0$  such that $\,\mathrm{dj}\!\left\{H_{\Phi}(\rho_n)\right\}\leq C\mathrm{dj}\!\left\{H(\rho_n)\right\}\,$  for any converging  sequence $\{\rho_n\}$ of input states;}
    \item \emph{the function $H_{\Phi}$ is continuous and
bounded on the set
$\,\mathrm{ext}\mathfrak{S}(\H_A)$;}\footnote{$\mathrm{ext}\mathfrak{S}(\H_A)$ is the set of pure states -- extreme points of the set $\mathfrak{S}(\H_A)$.}
    \item \emph{the function $H_{\Phi}$ is continuous on the cone
$\,\{\rho\in\mathfrak{T}_{+}(\H_A)\,|\,\mathrm{rank}\rho\leq
1\}$.}\smallskip
\end{enumerate}

\noindent\emph{If these properties hold then $\shs C=1$ in $\mathrm{(i)}$.}\medskip

\noindent B) \textit{Let $\,\Phi:A\rightarrow B$ be a quantum channel and $\,\widehat{\Phi}:A\rightarrow E$ its complementary channel defined by (\ref{c-channel}). Then for any converging sequence $\{\rho_n\}$ of input states
the following inequality holds
$$
\mathrm{dj}\!\left\{H_{\Phi}(\rho_n)\right\}\leq\mathrm{dj}\!\left\{H(\rho_n)\right\}+2\mathrm{dj}\!\left\{H_{\widehat{\Phi}}(\rho_n)\right\},
$$
where the factor $2$ can be removed if $\,\{H_{\widehat{\Phi}}(\rho_n)\}$ is a converging sequence.} \medskip

\noindent \textit{If
$\,\mathrm{dj}\!\left\{H_{\widehat{\Phi}}(\rho_n)\right\}\!=\!0$ (in particular, if $\,\dim E<+\infty$) then}
$$
\mathrm{dj}\!\left\{H_{\Phi}(\rho_n)\right\}\!=\!\mathrm{dj}\!\left\{H(\rho_n)\right\}.
$$
\end{theorem}

\begin{remark}\label{DB-OE-r}
By Theorem 2 in \cite{OE} properties $\mathrm{(ii)}$ and $\mathrm{(iii)}$ in Theorem \ref{DB-OE}A are equivalent
to preserving of local continuity of the entropy under action of the operation $\Phi$. So, quantum operations possessing these properties were called PCE\nobreakdash-\hspace{0pt}operations in \cite{OE}. The simplest examples of PCE\nobreakdash-\hspace{0pt}operations are quantum operations with finite Choi rank, for which property
$\mathrm{(iii)}$ in Theorem \ref{DB-OE} is directly
verified (since such operations have the Kraus representation with a finite number of summands).

If $\Phi$ is a \emph{channel} with finite Choi rank then
$\shs\mathrm{dj}\!\left\{H_{\Phi}(\rho_n)\right\}\!=\!\mathrm{dj}\!\left\{H(\rho_n)\right\}$ by Theorem \ref{DB-OE}B.
\end{remark}\medskip

\emph{Proof.} A) By Remark \ref{DB-OE-r} it suffices only to show that $\mathrm{(ii)}$ implies $\mathrm{(i)}$.

According to the general approximating technic used in the proof of Theorem 2 in \cite{OE} the functions $H$ and $H_{\Phi}$ are  pointwise limits of
the nondecreasing  sequences $\{H_{k}\}$ and  $\{H_{\Phi}^{k}\}$ of $k$-order approximators defined for any $\rho\in\S(\H_A)$ as follows
\begin{equation*}
H_{k}(\rho)=\sup_{\{\pi_{i},\rho_{i}\}\in\mathcal{P}_k(\rho)}
\sum_{i}\pi_{i}H(\rho_{i}),\quad H_{\Phi}^{k}(\rho)=\sup_{\{\pi_{i},\rho_{i}\}\in\mathcal{P}_k(\rho)}
\sum_{i}\pi_{i}H_{\Phi}(\rho_{i}),
\end{equation*}
where $\mathcal{P}_k(\rho)$ is the set of all countable ensembles with the average state $\rho$
consisting of states of rank $\leq k$. The functions $H_k$ are continuous on $\S(\H_A)$ for all $k$ by the strong stability of $\S(\H_A)$ \cite{SSP} while $\mathrm{(ii)}$ implies  continuity of all the functions $H_{\Phi}^k$ on $\S(\H_A)$ \cite{OE}.

By concavity of the function $\eta(x)=-x\log x$ and monotonicity of the relative entropy we have (cf.\cite{OE})
$$
\begin{array}{rl}
H_{\Phi}(\rho)-H_{\Phi}^k(\rho)\!\!&\,\leq\displaystyle\inf_{\{\pi_{i},\rho_{i}\}\in\mathcal{P}_k(\rho)}\sum_{i}\pi_{i}H(\Phi(\rho_{i})\|\Phi(\rho))
\\\\ &\leq\,
\displaystyle\inf_{\{\pi_{i},\rho_{i}\}\in\mathcal{P}_k(\rho)}\sum_{i}\pi_{i}H(\rho_{i}\|\rho)=H(\rho)-H_k(\rho),
\end{array}
$$
for any input state $\rho$ with finite $H(\rho)$. So, the validity of $\mathrm{(i)}$  with $C=1$ follows from Lemma \ref{app-l} in Section 2.

B) Since
\begin{equation}\label{ext-2}
H_{\Phi}(\rho)+H_{\widehat{\Phi}}(\rho)=H(\rho)+I(B\!:\!E)_{V\rho
V^*},\quad \rho\in\S(\H_A),
\end{equation}
where $V:A\rightarrow BE$
is any Stinespring isometry for $\Phi$, this assertion
follows from Theorem \ref{MI-DB} and the lower semicontinuity of all the terms in (\ref{ext-2}).  $\square$

\subsection{Information characteristics of a quantum channel}

In this section we will consider three basic characteristics of a quantum channel: the
constrained Holevo capacity, the quantum mutual information and the coherent information. We will obtain estimates for discontinuity jumps of these characteristics with respect to simultaneous variations of a channel and of an input state.

The \emph{constrained Holevo capacity} of a quantum channel
$\Phi:A\rightarrow B$  at a state $\rho\in\S(\H_A)$ is
defined as follows
\begin{equation*}
\bar{C}(\Phi,\rho)=\sup_{\sum_{i}\pi
_{i}\rho_{i}=\rho}\sum_{i}\pi_{i}H(\Phi (\rho
_{i})\|\Phi(\rho))=H(\Phi(\rho))-\inf_{\sum_{i}\pi
_{i}\rho_{i}=\rho}\sum_{i}\pi_{i}H(\Phi(\rho _{i})),
\end{equation*}
where the supremum (the infimum) is over all  countable
ensembles $\{\pi_i,\rho_i\}$ of input states with the average state $\rho$ and the
second formula is valid under the condition $H(\Phi(\rho))<+\infty$. This quantity plays a basic role in analysis of the classical capacity of a quantum channel (see details in \cite[Ch.8]{H-SCI}).\smallskip

The \emph{quantum mutual information} is an important characteristic of a
quantum channel related to its  entanglement-assisted classical
capacity \cite{H-SCI,N&Ch}. For a finite-dimensional
channel $\,\Phi:A\rightarrow B\,$ it can be defined as
\begin{equation*}
I(\Phi,\rho)=H(\rho)+H(\Phi(\rho))-H(\Phi,\rho),
\end{equation*}
where $H(\Phi,\rho)$ is the entropy exchange of the channel $\Phi$ at a state $\rho$ coinciding with the output entropy $H(\widehat{\Phi}(\rho))$ of any complementary channel $\widehat{\Phi}$ to the channel $\Phi$ (see Section 2).
In infinite dimensions this definition may contain the uncertainty
$"\infty-\infty"$, but it can be modified to avoid this problem as
follows
\begin{equation}\label{mi-qc+}
 I(\Phi,\rho) = H\left(\Phi \otimes \mathrm{Id}_{R}
(\hat{\rho})\shs \|\shs \Phi
(\rho) \otimes \varrho\shs\right),
\end{equation}
where $\hat{\rho}$ is a purification of the state $\rho$
in $\S(\mathcal{H}_{AR})$ and
$\varrho=\mathrm{Tr}_{A}\hat{\rho}$.
For an arbitrary quantum channel $\Phi$ the nonnegative function
$\rho\mapsto I( \Phi,\rho)$ defined by (\ref{mi-qc+}) is concave and
lower semicontinuous on the set $\mathfrak{S}(\mathcal{H}_A)$
\cite{H-SCI}. \smallskip

The \emph{coherent
information}
\begin{equation}\label{CI-rep+}
I_c(\Phi,\rho)\doteq
H(\Phi(\rho))-H(\Phi,\rho)
\end{equation}
of a channel $\Phi$ at a state $\rho$ is an important
characteristic related to the quantum capacity of a channel
\cite{H-SCI,N&Ch}. More suitable representation for the coherent information in infinite dimensions is given by the formula
\begin{equation}\label{CI-rep}
I_c(\Phi,\rho)=I(\Phi,\rho)-H(\rho),
\end{equation}
where $I(\Phi,\rho)$ is the mutual information defined by (\ref{mi-qc+}). This formula correctly determines a value in $[-H(\rho),H(\rho)]$ for any input state $\rho$ with finite entropy (despite possible infinite values of $H(\Phi(\rho))$ and $H(\Phi,\rho)=H(\widehat{\Phi}(\rho))$).

\smallskip

We will obtain estimates for discontinuity jumps of the above characteristic considered as functions of a pair $(\Phi,\rho)$, i.e. as functions on the Cartesian product of the set
$\mathfrak{F}_{AB}$ of all quantum channels from $A$ to $B$ equipped
with an appropriate topology (type of convergence) and the set
$\mathfrak{S}(\mathcal{H}_A)$ of input states. Such consideration is necessary for study of variation of quantum channel capacities with respect to variation of a channel and for analysis of quantum channels by approximation \cite{AQC}. We will assume that
the set $\mathfrak{F}_{AB}$ is equipped with the \emph{strong convergence topology} \cite{AQC}, in which convergence of a sequence
$\{\Phi_{n}\}\subset\mathfrak{F}_{AB}$ to a  channel
$\Phi_{0}\in\mathfrak{F}_{AB}$ means that
$$
\lim_{n\rightarrow+\infty}\Phi_{n}(\rho)=\Phi_{0}(\rho)\quad
\forall\rho\in\mathfrak{S}(\mathcal{H}_A).
$$
Preferability of using this topology in the infinite-dimensional case in comparison with the stronger topology induced by the norm of complete boundedness is discussed in Section 8.2. in \cite{CMI}. \medskip

The functions $(\Phi,\rho)\mapsto \bar{C}(\Phi,\rho)$  and $(\Phi,\rho)\mapsto I(\Phi,\rho)$ are lower semicontinuous on $\mathfrak{F}_{AB}\times\mathfrak{S}(\mathcal{H}_A)$ \cite{AQC}, so
discontinuity jumps of these functions for given sequences $\{\Phi_{n}\}\subset\mathfrak{F}_{AB}$ and
$\{\rho_n\}\subset\S(\H_A)$ converging respectively to a  channel
$\Phi_{0}\in\mathfrak{F}_{AB}$ and to a state $\rho_0\in\S(\H_A)$ are characterised by the nonnegative values
$$
\mathrm{dj}\!\left\{\bar{C}\!\left(\Phi_{n},\rho_n\right)\right\}\doteq\limsup_{n\rightarrow+\infty}\bar{C}(\Phi_n,\rho_n)-\bar{C}(\Phi_0,\rho_0)
$$
and
$$
\mathrm{dj}\!\left\{I\!\left(\Phi_{n},\rho_n\right)\right\}\doteq\limsup_{n\rightarrow+\infty}I(\Phi_n,\rho_n)-I(\Phi_0,\rho_0)
$$
(it is assumed that $\,\mathrm{dj}\!\left\{X\!\left(\Phi_{n},\rho_n\right)\right\}=+\infty\,$ if $\,X(\Phi_0,\rho_0)<+\infty$, $X=\bar{C},I$).

\smallskip

\begin{property}\label{IQ-DB} \emph{For any sequences $\{\Phi_{n}\}\subset\mathfrak{F}_{AB}$ and
$\{\rho_n\}\subset\S(\H_A)$ converging respectively to a  channel
$\Phi_{0}\in\mathfrak{F}_{AB}$ and to a state $\rho_0\in\S(\H_A)$ the following inequalities hold}
\begin{equation}\label{IQ-DB-1}
\mathrm{dj}\!\left\{\bar{C}\!\left(\Phi_{n},\rho_n\right)\right\} \leq \mathrm{dj}\!\left\{H(\Phi_n(\rho_n))\right\},\qquad\qquad\qquad\qquad\,
\end{equation}
\begin{equation}\label{IQ-DB-2}
\mathrm{dj}\!\left\{I\!\left(\Phi_{n},\rho_n\right)\right\} \leq 2\min\left\{\mathrm{dj}\!\left\{H(\rho_n)\right\},\mathrm{dj}\!\left\{H(\Phi_n(\rho_n))\right\}\right\}.
\end{equation}
\end{property}\smallskip

\emph{Proof.} Let $\overline{\mathrm{co}}H_{\Phi}$ be the convex closure of the output entropy of the channel $\Phi$
-- the maximal lower semicontinuous convex function on $\S(\H_A)$  not exceeding the function $H_{\Phi}=H(\Phi(\cdot))$.
Inequality (\ref{IQ-DB-1}) is proved by applying Lemma \ref{vsl+} to the identity
\begin{equation}\label{B-ident}
\bar{C}(\Phi,\rho)+\overline{\mathrm{co}}H_{\Phi}(\rho)=H(\Phi(\rho))
\end{equation}
valid for any $\,\rho\in\S(\H_A)$, and by using the lower semicontinuity of the function
$(\Phi,\rho)\mapsto\overline{\mathrm{co}}H_{\Phi}(\rho)$  on $\,\mathfrak{F}_{AB}\times\mathfrak{S}(\mathcal{H}_A)$ \cite{AQC}.\smallskip

Inequality (\ref{IQ-DB-2}) is proved by applying Theorem \ref{MI-DB} and Lemma \ref{p-lemma} to representation (\ref{mi-qc+}), since for any system $R$
the strong convergence of a sequence
$\{\Phi_n\}$ to a channel $\Phi_0$ implies the strong convergence of
the sequence $\{\Phi_n\otimes \id_{R}\}$ to the channel
$\Phi_0\otimes \id_{R}$. We have only to note that $H(\varrho)=H(\rho)$ for the state $\varrho$ in (\ref{mi-qc+}). $\square$ \smallskip

\begin{remark}\label{IQ-DB-r} By lower semicontinuity of the function
$(\Phi,\rho)\mapsto\bar{C}\!\left(\Phi,\rho\right)$ on $\,\mathfrak{F}_{AB}\times\mathfrak{S}(\mathcal{H}_A)$  identity (\ref{B-ident}) and Lemma \ref{vsl+} also imply
$$
\mathrm{dj}\!\left\{\overline{\mathrm{co}}H_{\Phi_n}(\rho_n)\right\} \leq \mathrm{dj}\!\left\{H(\Phi_n(\rho_n))\right\}
$$
for any sequences $\{\Phi_{n}\}\subset\mathfrak{F}_{AB}$ and
$\{\rho_n\}\subset\S(\H_A)$ converging respectively to a  channel
$\Phi_{0}\in\mathfrak{F}_{AB}$ and to a state $\rho_0\in\S(\H_A)$.
\end{remark}\medskip

The function $(\Phi,\rho)\mapsto I_c(\Phi,\rho)$ is defined by formula (\ref{CI-rep}) on the set $\,\mathfrak{F}_{AB}\times\mathfrak{S}_{\mathrm{f}}(\mathcal{H}_A)$, where  $\mathfrak{S}_{\mathrm{f}}(\mathcal{H}_A)$ is the set of input states with finite entropy. This function is not upper or lower semicontinuous.\footnote{By using the arguments from the proof of Corollary \ref{c-i-ls} in Section 3.2 one can show lower semicontinuity of the coherent information $I_c(\Phi,\rho)$ on the set of all pairs $\,(\Phi,\rho)$, where $\Phi$ is a pseudo-diagonal channel and $\rho$ is a state such that $H(\Phi(\rho))<+\infty$.} So, its discontinuity
for given sequences $\{\Phi_{n}\}\subset\mathfrak{F}_{AB}$ and
$\{\rho_n\}\subset\S_{\mathrm{f}}(\H_A)$ converging respectively to a  channel
$\Phi_{0}\in\mathfrak{F}_{AB}$ and to a state $\rho_0\in\S_{\mathrm{f}}(\H_A)$ can be characterised by two nonnegative values
$$
\mathrm{dj}^\downarrow\!\left\{I_c\left(\Phi_{n},\rho_n\right)\right\}\doteq\max\!\left\{\,\limsup_{n\rightarrow+\infty}I_c(\Phi_n,\rho_n)-I_c(\Phi_0,\rho_0), \;0\,\right\}
$$
and
$$
\mathrm{dj}^\uparrow\!\left\{I_c\left(\Phi_{n},\rho_n\right)\right\}\doteq\max\!\left\{\, I_c(\Phi_0,\rho_0)-\liminf_{n\rightarrow+\infty}I_c(\Phi_n,\rho_n),\;0\,\right\}
$$
describing respectively the maximal loss and the maximal gain of the coherent information corresponding to these sequences.\smallskip

\begin{remark}\label{CI-r} The function  $(\Phi,\rho)\mapsto H(\Phi,\rho)$ is lower semicontinuous on $\mathfrak{F}_{AB}\times\mathfrak{S}(\mathcal{H}_A)$. This follows from the representation $H(\Phi,\rho)=H\!\left(\Phi\otimes \mathrm{Id}_{R}
(\hat{\rho})\right)$, where $\hat{\rho}$ is a purification of the state $\rho$
in $\S(\mathcal{H}_{AR})$, and the arguments at the end of the proof of Proposition \ref{IQ-DB}.\footnote{We can not use the representation $H(\Phi,\rho)=H(\widehat{\Phi}(\rho))$, since in general strong convergence of a sequence  $\{\Phi_{n}\}\subset\mathfrak{F}_{AB}$ to a  channel
$\Phi_{0}\in\mathfrak{F}_{AB}$ does not imply existence of the corresponding sequence $\{\widehat{\Phi}_{n}\}$ of complementary channels strongly converging to a channel $\widehat{\Phi}_{0}$ complementary to the channel $\Phi_0$.}
\end{remark}\smallskip

\begin{corollary}\label{CI-DB} \emph{Let $\{\rho_n\}\subset\S(\H_A)$ be a sequence converging  to a state $\rho_0\in\S(\H_A)$ such that $H(\rho_n)<+\infty$ for all $\,n\geq0$. Then for any sequences $\{\Phi_{n}\}\subset\mathfrak{F}_{AB}$ converging to a  channel
$\Phi_{0}\in\mathfrak{F}_{AB}$ the following inequalities hold
\begin{equation*}
\mathrm{dj}^\downarrow\!\left\{I_c\left(\Phi_{n},\rho_n\right)\right\}\leq  \min\left\{2\mathrm{dj}\!\left\{H(\rho_n)\right\},\mathrm{dj}\!\left\{H(\Phi_n(\rho_n))\right\}\right\},
\end{equation*}
\begin{equation}\label{CI-DB-ub}
\mathrm{dj}^\uparrow\!\left\{I_c\left(\Phi_{n},\rho_n\right)\right\}\leq \min\left\{\mathrm{dj}\!\left\{H(\rho_n)\right\},\mathrm{dj}\!\left\{H(\Phi_n,\rho_n)\right\}\right\}\!.\quad
\end{equation}
If $\,\{H(\rho_n)\}$ is a converging sequence  then the factor $2$ in the first inequality can be removed.}
\end{corollary}\medskip

\emph{Proof.}  The inequalities (\ref{CI-DB-ub}) and
$\,\mathrm{dj}^\downarrow\!\left\{I_c\left(\Phi_{n},\rho_n\right)\right\}\leq \mathrm{dj}\!\left\{H(\Phi_n(\rho_n))\right\}\,$
are derived from (\ref{CI-rep+}) and (\ref{CI-rep}) by using Lemma \ref{vsl+} and the lower semicontinuity of the functions $(\Phi,\rho)\mapsto H(\Phi(\rho))$, $(\Phi,\rho)\mapsto I(\Phi,\rho)$ and $(\Phi,\rho)\mapsto H(\Phi,\rho)$ (see Remark \ref{CI-r}).

If $\,\mathrm{dj}^\downarrow\!\left\{I_c\left(\Phi_{n},\rho_n\right)\right\}>0\,$ then representation (\ref{CI-rep}) implies
$$
\mathrm{dj}^\downarrow\!\left\{I_c\left(\Phi_{n},\rho_n\right)\right\}\leq\!
\left[\limsup_{n\rightarrow+\infty}I(\Phi_n,\rho_n)-I(\Phi_0,\rho_0)\right]-
\left[\liminf_{n\rightarrow+\infty}H(\rho_n)-H(\rho_0)\right]\!.
$$
So, the inequality $\,\mathrm{dj}^\downarrow\!\left\{I_c\left(\Phi_{n},\rho_n\right)\right\}\leq 2\mathrm{dj}\!\left\{H(\rho_n)\right\}\,$ follows from Proposition \ref{IQ-DB} and the lower semicontinuity of the function $H(\rho)$. If $\{H(\rho_n)\}$ is a converging sequence  then  the second term in the above inequality coincides with
$\mathrm{dj}\!\left\{H(\rho_n)\right\}$. $\square$ \bigskip

I am grateful to A.S.Holevo and to the participants of his seminar
"Quantum probability, statistic, information" (the Steklov
Mathematical Institute) for useful discussion. I am also grateful to A.Winter for clarifying the particular questions concerning measures of classical correlations.


\end{document}